\documentclass[12pt]{article}
\usepackage{oldgerm}
\usepackage{yfonts}
\usepackage{euler}
\usepackage{amsmath}
\usepackage{amssymb}
\usepackage{amstext}
\usepackage{amscd}
\usepackage{amsfonts}
\DeclareMathAlphabet{\EuFrak}{U}{euf}{m}{n}
\DeclareMathAlphabet{\EuScript}{U}{eus}{m}{n}

\title{{\bf Convolution of Lorentz Invariant
Ultradistributions and Field Theory}
\thanks{\it{This work was partially supported by Consejo
Nacional
de Investigaciones Cient\'{\i}ficas and Comisi\'{o}n de
Investigaciones Cient\'{\i}ficas de la Pcia. de Buenos
Aires;
Argentina.}}}
\author{C.G.Bollini and M.C.Rocca\\
Departamento de F\'{\i}sica, Fac. de Ciencias Exactas,\\
Universidad Nacional de La Plata.\\
C.C. 67 (1900) La Plata. Argentina.}

\date{September 1, 2003}

\begin{document}

\maketitle

\begin{abstract}

In this work, a general definition of convolution between two
arbitrary four dimensional Lorentz invariant (fdLi) Tempered
Ultradistributions is given, in both: Minkowskian and Euclidean Space
(Spherically symmetric tempered ultradistributions).

The product of two arbitrary  fdLi distributions of
exponential type is defined via the convolution of its
corresponding Fourier Transforms.

Several examples of convolution of two fdLi Tempered
Ultradistributions are given. In particular we calculate
exactly the convolution of two Feynman's massless propagators.

An expression for the Fourier Transform of a Lorentz invariant
Tempered Ultradistribution in terms of modified Bessel distributions is obtained
in this work (Generalization of Bochner's formula to Minkowskian space).

At the same time, and in a previous step used for the deduction of
the convolution formula, we obtain the generalization to the
Minkowskian space, of the dimensional
regularization of the perturbation theory of Green Functions in
the Euclidean configuration space given in ref.[12]. As an example we
evaluate the convolution of two n-dimensional complex-mass Wheeler's
propagators.


\end{abstract}

\newpage

\renewcommand{\theequation}{\arabic{section}.\arabic{equation}}

\section{Introduction}

The question of the product of distributions with coincident point
singularities is related in Field Theory, to the asymptotic
behavior of loop integrals of propagators.

From a mathematical point of view, practically all definitions lead
to limitations on the set of distributions that can be multiplied
together to give another distribution of the same kind.

The properties of ultradistributions (ref.\cite{tp1,tp2}) are well
adapted for their use in Field Theory. In this respect we have
shown (ref.\cite{tp3}) that it is possible to define in one
dimensional space, the convolution of any pair of tempered
ultradistributions, giving as a result another tempered
ultradistribution. The next step is to consider the convolution
of any pair of tempered ultradistribution in n-dimensional space.
This follows from the formula obtained
in ref.\cite{tp3} for one dimensional space (See ref.\cite{tq1}.)

However, the resultant formula is rather complex  to be used
in practical applications and calculus. Then, for applications,
it is convenient to consider the convolution of any two
tempered ultradistributions which are even in the variables $k^0$ y $\rho$
(See ref.\cite{tq1}.).

A further step is to consider the convolution of two Lorentz invariant
tempered ultradistributions (See Section 7)

Ultradistributions also have the
advantage of being representable by means of analytic functions.
So that, in general, they are easier to work with them and,
as we shall see, have interesting properties. One of those properties
is that Schwartz tempered distributions are canonical and continuously
injected
into tempered ultradistributions and as a consequence the Rigged
Hilbert Space with tempered distributions is canonical and continuously
included
in the Rigged Hilbert Space with tempered ultradistributions.

This paper is organized as follow: in sections 2 and 3 we define
the Distributions of Exponential Type and the Fourier transformed
Tempered Ultradistributions. Each of them is part of a Guelfand's
Triplet ( or Rigged Hilbert Space \cite{tp4} ) together with their
respective duals and a ``middle term'' Hilbert space. In section 4
we give a general expression for the Fourier transform of a
spherically symmetric tempered ultradistributions and some
examples of it. In section 5 we obtain  the expression for the
Fourier transform of Lorentz invariant tempered ultradistributions
and we give some examples of its use. In section 6, we give the
general formula for the convolution of two spherically symmetric
tempered ultradistributions and followed by some examples. In
particular we evaluate exactly the convolution of two Feynman's
massless propagators. In section 7 we treat the convolution of two
Lorentz invariant tempered ultradistributions in Minkowskian
space. In subsection 1, we give the generalization to Minkowskian
space of the ``dimensional regularization in configuration space''
obtained in ref.\cite{tp12}. As an example of its use we evaluate
convolution of two complex mass Wheeler's propagators. In
subsection 2 we treat the central topic of this paper: the formula
for the convolution of two Lorentz invariant tempered
ultradistributions. Finally, section 8 is reserved for a
discussion of the principal results.

\section{Distributions of Exponential Type}

For the sake of the reader we shall present a brief description of the
principal properties of Tempered Ultradistributions.

{\bf Notations}.
The notations are almost textually taken from ref\cite{tp2}.
Let $\boldsymbol{{\mathbb{R}}^n}$
(res. $\boldsymbol{{\mathbb{C}}^n}$) be the real (resp. complex)
n-dimensional space whose points are denoted by $x=(x_1,x_2,...,x_n)$
(resp $z=(z_1,z_2,...,z_n)$). We shall use the notations:

(i) $x+y=(x_1+y_1,x_2+y_2,...,x_n+y_n)$\; ; \;
    $\alpha x=(\alpha x_1,\alpha x_2,...,\alpha x_n)$

(ii)$x\geqq 0$ means $x_1\geqq 0, x_2\geqq 0,...,x_n\geqq 0$

(iii)$x\cdot y=\sum\limits_{j=1}^n x_j y_j$

(iV)$\mid x\mid =\sum\limits_{j=1}^n \mid x_j\mid$

Let $\boldsymbol{{\mathbb{N}}^n}$ be the set of n-tuples of natural
numbers. If $p\in\boldsymbol{{\mathbb{N}}^n}$, then
$p=(p_1, p_2,...,p_n)$,
and $p_j$ is a natural number, $1\leqq j\leqq n$. $p+q$ denote
$(p_1+q_1, p_2+q_2,..., p_n+q_n)$ and $p\geqq q$ means $p_1\geqq q_1,
p_2\geqq q_2,...,p_n\geqq q_n$. $x^p$ means $x_1^{p_1}x_2^{p_2}...
x_n^{p_n}$. We shall denote by
$\mid p\mid=\sum\limits_{j=1}^n  p_j $ and by $D^p$ we denote the
differential operator ${\partial}^{p_1+p_2+...+p_n}/\partial{x_1}^{p_1}
\partial{x_2}^{p_2}...\partial{x_n}^{p_n}$

For any natural $k$ we define $x^k=x_1^k x_2^k...x_n^k$
and ${\partial}^k/\partial x^k=
{\partial}^{nk}/\partial x_1^k\partial x_2^k...\partial x_n^k$

The space $\boldsymbol{{\cal H}}$  of test functions
such that $e^{p|x|}|D^q\phi(x)|$ is bounded for any p and q
is defined ( ref.\cite{tp2} ) by means
of the countably set of norms:
\begin{equation}
\label{ep2.1}
{\|\hat{\phi}\|}_p=\sup_{0\leq q\leq p,\,x}
e^{p|x|} \left|D^q \hat{\phi} (x)\right|\;\;\;,\;\;\;p=0,1,2,...
\end{equation}
According to reference\cite{tp5} $\boldsymbol{{\cal H}}$  is a
$\boldsymbol{{\cal K}\{M_p\}}$ space
with:
\begin{equation}
\label{ep2.2}
M_p(x)=e^{(p-1)|x|}\;\;\;,\;\;\; p=1,2,...
\end{equation}
$\boldsymbol{{\cal K}\{e^{(p-1)|x|}\}}$ satisfies condition
$\boldsymbol({\cal N})$
of Guelfand ( ref.\cite{tp4} ). It is a countable Hilbert and nuclear
space:
\begin{equation}
\label{ep2.3}
\boldsymbol{{\cal K}\{e^{(p-1)|x|}\}} =\boldsymbol{{\cal H}} =
\bigcap\limits_{p=1}^{\infty}\boldsymbol{{\cal H}_p}
\end{equation}
where $\boldsymbol{{\cal H}_p}$ is obtained by completing
$\boldsymbol{{\cal H}}$ with the norm induced by
the scalar product:
\begin{equation}
\label{ep2.4}
{<\hat{\phi}, \hat{\psi}>}_p = \int\limits_{-\infty}^{\infty}
e^{2(p-1)|x|} \sum\limits_{q=0}^p D^q \overline{\hat{\phi}} (x) D^q
\hat{\psi} (x)\;dx \;\;\;;\;\;\;p=1,2,...
\end{equation}
where $dx=dx_1\;dx_2...dx_n$

If we take the usual scalar product:
\begin{equation}
\label{ep2.5}
<\hat{\phi}, \hat{\psi}> = \int\limits_{-\infty}^{\infty}
\overline{\hat{\phi}}(x) \hat{\psi}(x)\;dx
\end{equation}
then $\boldsymbol{{\cal H}}$, completed with (2.5), is the Hilbert space
$\boldsymbol{H}$
of square integrable functions.

The space of continuous linear functionals defined on
$\boldsymbol{{\cal H}}$ is the space
$\boldsymbol{{\Lambda}_{\infty}}$ of the distributions of the exponential
type ( ref.\cite{tp2} ).

The ``nested space''
\begin{equation}
\label{ep2.6}
\textgoth{\Large{H}}=
\boldsymbol{(}\boldsymbol{{\cal H}},\boldsymbol{H},
\boldsymbol{{\Lambda}_{\infty}} \boldsymbol{)}
\end{equation}
is a Guelfand's triplet ( or a Rigged Hilbert space \cite{tp4} ).

In addition we have: $\boldsymbol{{\cal H}}\subset\boldsymbol{{\cal S}}
\subset\boldsymbol{H}\subset\boldsymbol{{\cal S}^{'}}\subset
\boldsymbol{{\Lambda}_{\infty}}$, where $\boldsymbol{{\cal S}}$ is the
Schwartz space of rapidly decreasing test functions (ref\cite{tp6}).

Any Guelfand's triplet
$\textgoth{\Large{G}}=\boldsymbol{(}\boldsymbol{\Phi},
\boldsymbol{H},\boldsymbol{{\Phi}^{'}}\boldsymbol{)}$
has the fundamental property that a linear and symmetric operator
on $\boldsymbol{\Phi}$, admitting an extension to a self-adjoint
operator in
$\boldsymbol{H}$, has a complete set of generalized eigen-functions
in $\boldsymbol{{\Phi}^{'}}$ with real eigenvalues.

\section{Tempered Ultradistributions}
\setcounter{equation}{0}

The Fourier transform of a function $\hat{\phi}\in \boldsymbol{{\cal H}}$
is
\begin{equation}
\label{ep3.1}
\phi(z)=\frac {1} {2\pi}
\int\limits_{-\infty}^{\infty}\overline{\hat{\phi}}(x)\;e^{iz\cdot x}\;dx
\end{equation}
$\phi(z)$ is entire analytic and rapidly decreasing on straight lines
parallel
to the real axis. We shall call $\boldsymbol{{\EuFrak H}}$
the set of all such functions.
\begin{equation}
\label{ep3.2}
\boldsymbol{{\EuFrak H}}={\cal F}\left\{\boldsymbol{{\cal H}}\right\}
\end{equation}
It is a $\boldsymbol{{\cal Z}\{M_p\}}$ space ( ref.\cite{tp5} ),
countably normed and complete, with:
\begin{equation}
\label{ep3.3}
M_p(z)= (1+|z|)^p
\end{equation}
$\boldsymbol{{\EuFrak H}}$ is also a nuclear space with norms:
\begin{equation}
\label{ep3.4}
{\|\phi\|}_{pn} = \sup_{z\in V_n} {\left(1+|z|\right)}^p
|\phi (z)|
\end{equation}
where $V_k=\{z=(z_1,z_2,...,z_n)\in\boldsymbol{{\mathbb{C}}^n}:
\mid Im z_j\mid\leqq k, 1\leqq j \leqq n\}$

We can define the usual scalar product:
\begin{equation}
\label{ep3.5}
<\phi (z), \psi (z)>=\int\limits_{-\infty}^{\infty}
\phi(z) {\psi}_1(z)\;dz =
\int\limits_{-\infty}^{\infty} \overline{\hat{\phi}}(x)
\hat{\psi}(x)\;dx
\end{equation}
where:
\[{\psi}_1(z)=\int\limits_{-\infty}^{\infty}
\hat{\psi}(x)\; e^{-iz\cdot x}\;dx\]
and $dz=dz_1\;dz_2...dz_n$

By completing $\boldsymbol{{\EuFrak H}}$ with the norm induced by (3.5)
we get the Hilbert space of square integrable functions.

The dual of $\boldsymbol{{\EuFrak H}}$ is the space
$\boldsymbol{{\cal U}}$ of tempered ultradistributions
( ref.\cite{tp2} ). In other words, a tempered ultradistribution is
a continuous linear functional defined on the space
$\boldsymbol{{\EuFrak H}}$ of entire
functions rapidly decreasing on straight lines parallel to the real axis.

The set
$\textgoth{\Large{U}}=
\boldsymbol{({\EuFrak H},H,{\cal U})}$ is also a Guelfand's triplet.

Moreover, we have: $\boldsymbol{{\EuFrak H}}\subset\boldsymbol{{\cal S}}
\subset\boldsymbol{H}\subset\boldsymbol{{\cal S}^{'}}\subset
\boldsymbol{{\cal U}}$.

$\boldsymbol{{\cal U}}$ can also be characterized in the following way
( ref.\cite{tp2} ): let $\boldsymbol{{\cal A}_{\omega}}$ be the space of
all functions $F(z)$ such that:

${\Large {\boldsymbol{I}}}$-
$F(z)$ is analytic for $\{z\in \boldsymbol{{\mathbb{C}}^n} :
|Im(z_1)|>p, |Im(z_2)|>p,...,|Im(z_n)|>p\}$.

${\Large {\boldsymbol{II}}}$-
$F(z)/z^p$ is bounded continuous  in
$\{z\in \boldsymbol{{\mathbb{C}}^n} :|Im(z_1)|\geqq p,|Im(z_2)|\geqq p,
...,|Im(z_n)|\geqq p\}$,
where $p=0,1,2,...$ depends on $F(z)$.

Let $\boldsymbol{\Pi}$ be the set of all $z$-dependent pseudo-polynomials,
$z\in \boldsymbol{{\mathbb{C}}^n}$.
Then $\boldsymbol{{\cal U}}$ is the quotient space:

${\Large {\boldsymbol{III}}}$-
$\boldsymbol{{\cal U}}=\boldsymbol{{\cal A}_{\omega}/\Pi}$

By a pseudo-polynomial we understand a function of $z$ of the form $\;\;$
$\sum_s z_j^s G(z_1,...,z_{j-1},z_{j+1},...,z_n)$ with
$G(z_1,...,z_{j-1},z_{j+1},...,z_n)\in\boldsymbol{{\cal A}_{\omega}}$

Due to these properties it is possible to represent any ultradistribution
as ( ref.\cite{tp2} ):
\begin{equation}
\label{ep3.6}
F(\phi)=<F(z), \phi(z)>=\oint\limits_{\Gamma} F(z) \phi(z)\;dz
\end{equation}
$\Gamma={\Gamma}_1\cup{\Gamma}_2\cup ...{\Gamma}_n$
where the path ${\Gamma}_j$ runs parallel to the real axis from
$-\infty$ to $\infty$ for $Im(z_j)>\zeta$, $\zeta>p$ and back from
$\infty$ to $-\infty$ for $Im(z_j)<-\zeta$, $-\zeta<-p$.
( $\Gamma$ surrounds all the singularities of $F(z)$ ).

Formula (3.6) will be our fundamental representation for a tempered
ultradistribution. Sometimes use will be made of ``Dirac formula''
for ultradistributions ( ref.\cite{tp1} ):
\begin{equation}
\label{ep3.7}
F(z)=\frac {1} {(2\pi i)^n}\int\limits_{-\infty}^{\infty}
\frac {f(t)} {(t_1-z_1)(t_2-z_2)...(t_n-z_n)}\;dt
\end{equation}
where the ``density'' $f(t)$ is such that
\begin{equation}
\label{ep3.8}
\oint\limits_{\Gamma} F(z) \phi(z)\;dz =
\int\limits_{-\infty}^{\infty} f(t) \phi(t)\;dt
\end{equation}
While $F(z)$ is analytic on $\Gamma$, the density $f(t)$ is in
general singular, so that the r.h.s. of (3.8) should be interpreted
in the sense of distribution theory.

Another important property of the analytic representation is the fact
that on $\Gamma$, $F(z)$ is bounded by a power of $z$ ( ref.\cite{tp2} ):
\begin{equation}
\label{ep3.9}
|F(z)|\leq C|z|^p
\end{equation}
where $C$ and $p$ depend on $F$.

The representation (3.6) implies that the addition of a
pseudo-polynomial $P(z)$ to $F(z)$ do not alter the ultradistribution:
\[\oint\limits_{\Gamma}\{F(z)+P(z)\}\phi(z)\;dz=
\oint\limits_{\Gamma} F(z)\phi(z)\;dz+\oint\limits_{\Gamma}
P(z)\phi(z)\;dz\]
But:
\[\oint\limits_{\Gamma} P(z)\phi(z)\;dz=0\]
as $P(z)\phi(z)$ is entire analytic in some of the variables $z_j$
( and rapidly decreasing ),
\begin{equation}
\label{ep3.10}
\therefore \;\;\;\;\oint\limits_{\Gamma} \{F(z)+P(z)\}\phi(z)\;dz=
\oint\limits_{\Gamma} F(z)\phi(z)\;dz
\end{equation}

\section{The Fourier Transform in Euclidean Space}

\setcounter{equation}{0}

The Fourier transform of a spherically symmetric function $\hat{f}\in \boldsymbol{H}$ is given,
according to Bochner's formula by:
\begin{equation}
\label{ep4.1}
f(k)=\frac {(2\pi)^{\frac {\nu} {2}}} {k^{\frac {\nu-2} {2}}}\int\limits_0^{\infty}\hat{f}(r)
r^{\frac {\nu} {2}}\boldsymbol{\cal{J}}_{\frac {\nu-2} {2}}(kr)\;dr
\end{equation}
where $r=x_0^2+x_1^2+\cdot\cdot\cdot+x_{\nu-1}^2$\; ; \;
$k=k_0^2+k_1^2+\cdot\cdot\cdot+k_{\nu-1}^2$ and
$\boldsymbol{\cal{J}}_{\frac {\nu} {2}}$ is the
Bessel function of order $\nu-2 /2$.
By the use of the equality
\begin{equation}
\label{ep4.2}
\pi\boldsymbol{\cal{J}}_{\frac {\nu-2} {2}}(z)=
e^{-i\frac {\pi} {4}\nu}\boldsymbol{\cal{K}}_{\frac {\nu-2} {2}}(-iz)+
e^{i\frac {\pi} {4}\nu}\boldsymbol{\cal{K}}_{\frac {\nu-2} {2}}(iz)
\end{equation}
where $\boldsymbol{\cal{K}}$ is the modified Bessel function,
(\ref{ep4.1}) takes the form:
\[f(k)=2\frac {(2\pi)^{\frac {\nu-2} {2}}} {k^{\frac {\nu-2} {2}}}\int\limits_0^{\infty}\hat{f}(r)
r^{\frac {\nu} {2}}\left[
e^{-i\frac {\pi} {4}\nu}\boldsymbol{\cal{K}}_{\frac {\nu-2} {2}}(-ikr)+\right.\]
\begin{equation}
\label{ep4.3}
\left. e^{i\frac {\pi} {4}\nu}\boldsymbol{\cal{K}}_{\frac {\nu-2} {2}}(ikr)\right]\;dr
\end{equation}
By performing the change of variables $x=r^{\frac {1} {2}}$,$\rho=k^{\frac {1} {2}}$
(\ref{ep4.1}), and (\ref{ep4.3})can be re-written as:
\begin{equation}
\label{ep4.4}
f(\rho)=\pi\frac {(2\pi)^{\frac {\nu-2} {2}}} {\rho^{\frac {\nu-2} {4}}}\int\limits_0^{\infty}\hat{f}(x)
x^{\frac {\nu-2} {4}}\boldsymbol{\cal{J}}_{\frac {\nu-2} {2}}({\rho}^{1/2}x^{1/2})\;dx
\end{equation}
\[f(\rho)=\frac {(2\pi)^{\frac {\nu-2} {2}}} {\rho^{\frac {\nu-2} {4}}}\int\limits_0^{\infty}\hat{f}(x)
x^{\frac {\nu-2} {4}}\left[
e^{-i\frac {\pi} {4}\nu}\boldsymbol{\cal{K}}_{\frac {\nu-2} {2}}(-ix^{1/2}{\rho}^{1/2})+\right.\]
\begin{equation}
\label{ep4.5}
\left. e^{i\frac {\pi} {4}\nu}\boldsymbol{\cal{K}}_{\frac {\nu-2} {2}}(ix^{1/2}{\rho}^{1/2})\right]\;dx
\end{equation}
Here we have taken $\rho=\gamma+i\sigma$ and
\begin{equation}
\label{ep4.6}
{\rho}^{1/2}=\sqrt{\frac {\gamma+\sqrt{{\gamma}^2+{\sigma}^2}} {2}}+iSgn(\sigma)
\sqrt{\frac {-\gamma+\sqrt{{\gamma}^2+{\sigma}^2}} {2}}
\end{equation}
We can  extend (\ref{ep4.4}) to the
complex plane and obtain the corresponding ultradistribution. As a first step we calculate
the Fourier antitrasfom of
${\rho}^{\frac {2-\nu} {4}} {\cal{J}}_{\frac {\nu-2} {2}}(x^{1/2}{\rho}^{1/2})$.
We have:
\[\frac {1} {2\pi}\int\limits_0^{\infty}
{\rho}^{\frac {2-\nu} {4}} {\cal{J}}_{\frac {\nu-2} {2}}(x^{1/2}{\rho}^{1/2})
e^{-i\rho t}\;d\rho=\]
\begin{equation}
\label{ep4.7}
\frac {e^{\frac {i\pi(\nu-4)} {8}}(t-i0)^{\frac {\nu-4} {4}}}
{\pi x^{1/2}\Gamma(\frac {\nu} {2})}e^{\frac {ix} {8t}}
{\cal{M}}_{\frac {4-\nu} {4},\frac {\nu-2} {4}}\left(-\frac {ix} {4t}\right)
\end{equation}
We have used $\boldsymbol{6.631}.(1)$ of ref.\cite{tp8}
($\cal{M}$ is the Whittaker function).
Now we can use $\boldsymbol{9.233},(1),(2)$ of ref.\cite{tp8} and write
\[{\cal{M}}_{\frac {4-\nu} {4},\frac {\nu-2} {4}}\left(-\frac {ix} {4t}\right)=
\frac {\Gamma(\frac {\nu} {2})} {\Gamma(\frac {\nu-2} {2})}
e^{\frac {i\pi(4-\nu)} {4}}
{\cal{W}}_{\frac {\nu-4} {4},\frac {\nu-2} {4}}\left(\frac {ix} {4t}\right)+\]
\[\Gamma\left(\frac {\nu} {2}\right)e^{\frac {i\pi(2-\nu)} {2}}
{\cal{W}}_{\frac {4-\nu} {4},\frac {\nu-2} {4}}\left(-\frac {ix} {4t}\right)
\;\;\;\;t>0.\]
\[{\cal{M}}_{\frac {4-\nu} {4},\frac {\nu-2} {4}}\left(-\frac {ix} {4t}\right)=
\frac {\Gamma(\frac {\nu} {2})} {\Gamma(\frac {\nu-2} {2})}
e^{\frac {i\pi(\nu-4)} {4}}
{\cal{W}}_{\frac {\nu-4} {4},\frac {\nu-2} {4}}\left(\frac {ix} {4t}\right)+\]
\begin{equation}
\label{ep4.8}
\Gamma\left(\frac {\nu} {2}\right)e^{\frac {i\pi(\nu-2)} {2}}
{\cal{W}}_{\frac {4-\nu} {4},\frac {\nu-2} {4}}\left(-\frac {ix} {4t}\right)
\;\;\;\;t<0.
\end{equation}
As a second step we calculate the complex Fourier transform of the second term
of (\ref{ep4.7}) using (\ref{ep4.8}). We obtain:
\[{\cal{F}}_c\left[\frac {e^{\frac {i\pi(\nu-4)} {8}}(t-i0)^{\frac {\nu-4} {4}}}
{\pi x^{1/2}\Gamma(\frac {\nu} {2})}e^{\frac {ix} {8t}}
{\cal{M}}_{\frac {4-\nu} {4},\frac {\nu-2} {4}}\left(-\frac {ix} {4t}\right)\right](\rho)=\]
\[{\rho}^{\frac {2-\nu} {4}}\left\{\Theta[\Im(\rho)]e^{-\frac {i\pi\nu} {4}}
{\cal{K}}_{\frac {\nu-2} {2}}(-ix^{1/2}{\rho}^{1/2})-
\Theta[-\Im(\rho)]e^{\frac {i\pi\nu} {4}}
{\cal{K}}_{\frac {\nu-2} {2}}(ix^{1/2}{\rho}^{1/2})\right.+\]
\begin{equation}
\label{ep4.9}
\left.\frac {2^{\frac {4-\nu} {2}}i} {\Gamma(\frac {\nu-2} {2})}
{\cal{S}}_{\frac {\nu-4} {2},\frac {\nu-2} {2}}(x^{1/2}{\rho}^{1/2})\right\}
\end{equation}
where we have used $\boldsymbol{7.629},(1),(2)$ of ref.\cite{tp8} and
$\cal{S}$ is the Lommel function (ref.\cite{tp9}, pag 349, formula 3).
The corresponding ultradistribution is then defined as:
\[F(\rho)=\frac {(2\pi)^{\frac {\nu-2} {2}}} {{\rho}^{\frac {\nu-2} {4}}}
\int\limits_0^{\infty}\hat{f}(x)x^{\frac {\nu-2} {4}}\left\{\Theta[\Im(\rho)]
e^{-\frac {i\pi\nu} {4}}{\cal{K}}_{\frac {\nu-2} {2}}(-ix^{1/2}{\rho}^{1/2})-\right.\]
\[\left.\Theta[-\Im(\rho)]
e^{\frac {i\pi\nu} {4}}{\cal{K}}_{\frac {\nu-2} {2}}(ix^{1/2}{\rho}^{1/2})\right\}\;dx\; +\]
\begin{equation}
\label{ep4.10}
\frac {2{\pi}^{\frac {\nu-2} {2}}} {\Gamma(\frac {\nu-2} {2})
{\rho}^{\frac {\nu-2} {4}}}
\int\limits_0^{\infty}\hat{f}(x)x^{\frac {\nu-2} {4}} {\cal{S}}_{\frac {\nu-4} {2},
\frac {\nu-2} {2}}(x^{1/2}{\rho}^{1/2})\;dx
\end{equation}
When $\nu=2n$, n an entire number,
${\rho}^{\frac {2-\nu} {4}} {\cal {S}}_{\frac {\nu-4} {2},\frac {\nu-2} {2}}$
is equivalent to zero. In fact
\begin{equation}
\label{ep4.11}
{\rho}^{\frac {2-\nu} {4}} {\cal {S}}_{\frac {\nu-4} {2},\frac {\nu-2} {2}}=
\sum\limits_{m=0}^{\frac {\nu-4} {2}}\frac {(\frac {\nu} {2}-m)!} {m!}
4^{\frac {\nu-2-4m} {4}} x^{\frac {4m+2-\nu} {4}}{\rho}^{\frac {2m+2-\nu} {2}}
\end{equation}
(\ref{ep4.11}) is a polynomial in ${\rho}^{-1}$. However when the volume
element is taken into account that expression is transformed into a
polynomial in $\rho$ which according to (\ref{ep3.10}) is a null
ultradistribution.
Thus in this case the second integral in (\ref{ep4.10}) vanishes and it becomes in:
\[F(\rho)=\frac {(2\pi)^{\frac {\nu-2} {2}}} {\rho^{\frac {\nu-2} {4}}}\int\limits_0^{\infty}\hat{f}(x)
x^{\frac {\nu-2} {4}}\left[\Theta[\Im(\rho)]
e^{-i\frac {\pi} {4}\nu}\boldsymbol{\cal{K}}_{\frac {\nu-2} {2}}(-ix^{1/2}{\rho}^{1/2})\right.\]
\begin{equation}
\label{ep4.12}
\left. -\Theta[-\Im(\rho)]
 e^{i\frac {\pi} {4}\nu}\boldsymbol{\cal{K}}_{\frac {\nu-2} {2}}(ix^{1/2}{\rho}^{1/2})\right]\;dx
\end{equation}
Note that the complex Fourier transform (\ref{ep4.12}) is not merely the Fourier
transform (\ref{ep4.5}) in which the variable $\rho$ is considered to be a
complex number. (\ref{ep4.12}) gives the ultradistribution associated to $f(\rho)$.
In the next section we shall see that formulae  (\ref{ep4.5}), (\ref{ep4.12})
can be generalized to Minkowskian space.

When $\hat{f}$ is a spherically symmetric distribution of exponential type, we can use
(\ref{ep4.10}) to define its Fourier transform. In addition we can follow
the treatment of ref.\cite{tp10} to define the Fourier transform. Thus we have
\begin{equation}
\label{ep4.13}
\int\limits_0^{\infty}f(\rho)\phi(\rho){\rho}^{\frac {\nu-2} {2}}\;d\rho=(2\pi)^{\nu}
\int\limits_0^{\infty}\hat{f}(x)\hat{\phi}(x)x^{\frac {\nu-2} {2}}\;dx
\end{equation}
The corresponding tempered ultradistribution in the one-dimensional  complex
variable $\rho$ is obtained in the following way: let $\hat{g}(t)$ be defined as:
\begin{equation}
\label{ep4.14}
\hat{g}(t)=\frac {1} {(2\pi)^{\nu}}\int\limits_0^{\infty}f(\rho)e^{-i\rho t}\;d\rho
\end{equation}
Then:
\begin{equation}
\label{ep4.15}
F(\rho)=\Theta[\Im(\rho)]\int\limits_0^{\infty}\hat{g}(t)e^{i\rho t}\;dt-
\Theta[-\Im(\rho)]\int\limits_{-\infty}^0\hat{g}(t)e^{i\rho t}\;dt
\end{equation}
or if we use Dirac's formula
\begin{equation}
\label{ep4.16}
F(\rho)=\frac {1} {2\pi i}\int\limits_0^{\infty}\frac {f(t)} {t-\rho}\;dt
\end{equation}

The inversion formula $(\nu=2n)$ for $F(\rho)$ is given by
\begin{equation}
\label{ep4.17}
\hat{f}(x)=\frac {\pi} {(2\pi)^{\frac {\nu+2} {2}} x^{\frac {\nu-2} {4}}}
\oint\limits_{\Gamma} F(\rho) {\rho}^{\frac {\nu-2} {4}} {\cal J}_{\frac {\nu-2} {2}}
(x^{1/2}{\rho}^{1/2})\;d\rho
\end{equation}
Note that the factor multiplying
$F(\rho)$ is an entire function of $\rho$ for $\nu=2n$. In this case
the first term of (\ref{ep4.13}) takes the form:
\begin{equation}
\label{ep4.18}
\oint\limits_{\Gamma}F(\rho)\phi(\rho){\rho}^{\frac {\nu-2} {2}}\;d\rho=
(2\pi)^{\nu}\int\limits_0^{\infty}\hat{f}(x)\hat{\phi}(x)x^{\frac {\nu-2} {2}}\;dx
\end{equation}
We can now define a spherically symmetric tempered
ultradistribution as the complex Fourier transform
of a spherically symmetric
distribution of exponential type. Note that a spherically symmetric
ultradistribution is not  necessarily spherically symmetric in an explicit way.

We give now same examples of the use of Fourier transform.

\subsection*{Examples}

As a first example we calculate the complex Fourier transform of $e^{ar}$
(where a is a complex number) for $\nu=2n$. From (\ref{ep4.12}) we write:
\[F(\rho)=\frac {(2\pi)^{\frac {\nu-2} {2}}} {{\rho}^{\frac {\nu-2} {4}}}
\int\limits_0^{\infty}e^{ax^{1/2}}x^{\frac {\nu-2} {4}}\left\{
\Theta[\Im(\rho)]e^{-\frac {i\pi\nu} {4}}{\cal{K}}_{\frac {\nu-2} {2}}
(-ix^{1/2}{\rho}^{1/2})-\right.\]
\begin{equation}
\label{ep4.19}
\left.\Theta[-\Im(\rho)]e^{\frac {i\pi\nu} {4}}{\cal{K}}_{\frac {\nu-2} {2}}
(ix^{1/2}{\rho}^{1/2})\;dx\right\}
\end{equation}
Now:
\[\int\limits_0^{\infty}e^{ax^{1/2}}x^{\frac {\nu-2} {4}}
{\cal{K}}_{\frac {\nu-2} {2}} (-ix^{1/2}{\rho}^{1/2})=
2\sqrt{\pi}\;e^{\frac {i\pi(\nu+2)} {4}}
\frac {\Gamma(\nu)} {\Gamma(\frac {\nu+3} {2})} \frac {{\rho}^{\frac {\nu-2} {4}}}
{({\rho}^{1/2}-ia)}\;\times\]
\[\boldsymbol{F}\left(\nu ,\frac {\nu-1} {2} ,\frac {\nu+3} {2} ,\frac {a-i{\rho}^{1/2}} {a+i{\rho}^{1/2}}
\right)\,\,\,\,\,\Im(\rho)>0\]
\[\int\limits_0^{\infty}e^{ax^{1/2}}x^{\frac {\nu-2} {4}}
{\cal{K}}_{\frac {\nu-2} {2}} (ix^{1/2}{\rho}^{1/2})=
2\sqrt{\pi}\;e^{-\frac {i\pi(\nu+2)} {4}}
\frac {\Gamma(\nu)} {\Gamma(\frac {\nu+3} {2})} \frac {{\rho}^{\frac {\nu-2} {4}}}
{({\rho}^{1/2}+ia)}\;\times\]
\begin{equation}
\label{ep4.20}
\boldsymbol{F}\left(\nu ,\frac {\nu-1} {2} ,\frac {\nu+3} {2} ,\frac {a+i{\rho}^{1/2}} {a-i{\rho}^{1/2}}
\right)\,\,\,\,\,\Im(\rho)<0
\end{equation}
To obtain (\ref{ep4.20}) we have used $\boldsymbol{6.621}, (3)$ of ref.\cite{tp8}
(Here $\boldsymbol{F}$ is the hypergeometric function).
Then we have:
\[F(\rho)=(4\pi)^{\frac {\nu-2} {2}}i\frac {\Gamma(\nu)} {\Gamma(\frac {\nu+3} {2} )}
\left\{\frac {\Theta[\Im(\rho)]} {({\rho}^{1/2}-ia)}
\boldsymbol{F}\left(\nu ,\frac {\nu-1} {2} ,\frac {\nu+3} {2} ,\frac {a-i{\rho}^{1/2}} {a+i{\rho}^{1/2}}
\right)\;+\right.\]
\begin{equation}
\label{ep4.21}
\left.\frac {\Theta[-\Im(\rho)]} {({\rho}^{1/2}+ia)}
\boldsymbol{F}\left(\nu ,\frac {\nu-1} {2} ,\frac {\nu+3} {2} ,\frac {a+i{\rho}^{1/2}} {a-i{\rho}^{1/2}}
\right)\right\}
\end{equation}
As a second example we evaluate the Fourier antitransform of
$[-2\pi i(\rho-{\mu}^2)]^{-1}$ where $\mu$ is a complex number
and $\nu=2n$. Using (\ref{ep4.17}) we have:
\[\hat{f}(x)=-\frac {\pi} {(2\pi)^{\frac {\nu+2} {2}} x^{\frac {\nu-2} {4}}}
\oint\limits_{\Gamma}
\frac {{\rho}^{\frac {\nu-2} {4}}} {2\pi i(\rho-{\mu}^2)}
{\cal J}_{\frac {\nu-2} {2}}
(x^{1/2}{\rho}^{1/2})\;d\rho =\]
\begin{equation}
\label{ep4.22}
\frac {\pi{\mu}^{\frac {\nu-2} {2}}} {(2\pi)^{\frac {\nu+2} {2}}}
x^{\frac {2-\nu} {4}}{\cal{J}}_{\frac {\nu-2} {2}}(\mu x^{1/2})
\end{equation}
We can test the result (\ref{ep4.22}) by transforming it. Taking into account
that for $\nu$ even
${\cal{J}}_{\frac {\nu-2} {2}}=e^{\frac {i\pi(\nu-2)} {2}}
{\cal{J}}_{\frac {2-\nu} {2}}$. Thus:
\[F(\rho)=\frac {{\mu}^{\frac {\nu-2} {2}}} {4\pi}e^{\frac {i\pi(\nu-2)} {2}}
{\rho}^{\frac {2-\nu} {4}}\int\limits_0^{\infty}
{\cal{J}}_{\frac {2-\nu} {2}}(\mu x^{1/2})
\left\{\Theta[\Im(\rho)]e^{-\frac {i\pi\nu} {4}}
{\cal{K}}_{\frac {\nu-2} {2}}(-ix^{1/2}{\rho}^{1/2})\;-\right.\]
\begin{equation}
\label{ep4.23}
\left.\Theta[-\Im(\rho)]e^{\frac {i\pi\nu} {4}}
{\cal{K}}_{\frac {\nu-2} {2}}(ix^{1/2}{\rho}^{1/2})\right\}\;dx
\end{equation}
Now:
\[\int\limits_0^{\infty}{\cal{J}}_{\frac {2-\nu} {2}}(\mu x^{1/2})
{\cal{K}}_{\frac {\nu-2} {2}}(-ix^{1/2}{\rho}^{1/2})\;dx=
e^{\frac {i\pi(6-\nu)} {4}}{\mu}^{\frac {2-\nu} {2}}
\frac {{\rho}^{\frac {\nu-2} {4}}} {\rho-{\mu}^2}\;;\;\Im(\rho)>0\]
\begin{equation}
\label{ep4.24}
\int\limits_0^{\infty}{\cal{J}}_{\frac {2-\nu} {2}}(\mu x^{1/2})
{\cal{K}}_{\frac {\nu-2} {2}}(ix^{1/2}{\rho}^{1/2})\;dx=
e^{-\frac {i\pi(6-\nu)} {4}}{\mu}^{\frac {2-\nu} {2}}
\frac {{\rho}^{\frac {\nu-2} {4}}} {\rho-{\mu}^2}\;;\;\Im(\rho)<0
\end{equation}
where we have used $\boldsymbol{6.576}, (3)$ of ref.\cite{tp8}. Then we have:
\begin{equation}
\label{ep4.25}
F(\rho)=-\frac {1} {2\pi i (\rho-{\mu}^2)}
\end{equation}
As a third example we give the Fourier transform of $\delta(x-a)$
for all $\nu$. Using (\ref{ep4.10}) we obtain:
\[F(\rho)=\frac {(2\pi)^{\frac {\nu-2} {2}}} {{\rho}^{\frac {\nu-2} {4}}}
a^{\frac {\nu-2} {4}}\left\{\Theta[\Im(\rho)]
e^{-\frac {i\pi\nu} {4}}{\cal{K}}_{\frac {\nu-2} {2}}(-ia^{1/2}{\rho}^{1/2})-\right.\]
\[\left.\Theta[-\Im(\rho)]
e^{\frac {i\pi\nu} {4}}{\cal{K}}_{\frac {\nu-2} {2}}(ia^{1/2}{\rho}^{1/2})\right\}\;\; +\]
\begin{equation}
\label{ep4.26}
\frac {2{\pi}^{\frac {\nu-2} {2}}} {\Gamma(\frac {\nu-2} {2})
{\rho}^{\frac {\nu-2} {4}}}
a^{\frac {\nu-2} {4}} {\cal{S}}_{\frac {\nu-4} {2},
\frac {\nu-2} {2}}(a^{1/2}{\rho}^{1/2})
\end{equation}
The reader can verify that the cut of (\ref{ep4.26}) along the negative real axis is zero.

\section{The Fourier Transform in Minkowskian Space}

\setcounter{equation}{0}

For the Minkowskian case we begin with the formula:

\begin{equation}
\label{ep5.1}
f(k_0,k)=\frac {(2\pi)^{\frac {\nu-1} {2}}} {k^{\frac {\nu-3} {2}}}
\int\limits_{-\infty}^{\infty}\int\limits_0^{\infty}
\hat{f}(x_0,r)r^{\frac {\nu-1} {2}}{\cal{J}}_{\frac {\nu-3} {2}} (kr)
e^{ik_0x^0}\;dx^0\;dr
\end{equation}
that can be re-written as:
\[f(k_0^2-k^2)=\frac {(2\pi)^{\frac {\nu-3} {2}}} {k^{\frac {\nu-3} {2}}}
\int\limits_{-\infty}^{\;\;\;\infty}\int\limits_{-\infty}^{\;\;\;\infty}
\int\limits_{-\infty}^{\;\;\;\infty}\int\limits_0^{\infty}
\hat{f}(x)e^{it(x-s_0^2+s^2)}
s^{\frac {\nu-1} {2}}{\cal{J}}_{\frac {\nu-3} {2}} (ks)\;\times\]
\begin{equation}
\label{ep5.2}
e^{ik_0s^0}\;dt\;dx\;ds^0\;ds
\end{equation}
Now:
\begin{equation}
\label{ep5.3}
\int\limits_0^{\infty}e^{its^2}s^{\frac {\nu-1} {2}}{\cal{J}}_{\frac {\nu-3} {2}}(ks)\;ds=
\frac {1} {2} {\left(\frac {k} {2}\right)}^{\frac {\nu-3} {2}}
(t+i0)^{\frac {1-\nu} {2}}
e^{i\left[\frac {\pi} {2} (\frac {\nu-1} {2})-\frac {k^2} {4t}\right]}
\end{equation}
\begin{equation}
\label{ep5.4}
\int\limits_{-\infty}^{\infty}e^{-its_0^2}e^{ik_0s^0}\;ds^0=
\sqrt{\pi} (t-i0)^{-\frac {1} {2}} e^{i\left(\frac {k_0^2} {4t}-\frac {\pi} {4}\right)}
\end{equation}
We have used $\boldsymbol{6.631}, (4)$ and
$\boldsymbol{3.462},(3)$ of ref.\cite{tp8}.
Then we obtain for (\ref{ep5.2}):
With the results (\ref{ep5.3}),(\ref{ep5.4}) we obtain for (\ref{ep5.2}):
\[f(k_0^2-k^2)=\frac {(2\pi)^{\frac {\nu-3} {2}}} {2^{\frac {\nu-1} {2}}}
\sqrt{\pi}\; e^{\frac {i\pi(\nu-2)} {4}}
\int\limits_{-\infty}^{\infty}\int\limits_0^{\infty}\hat{f}(x)
\left[e^{itx}e^{\frac {i(k_0^2-k^2)} {4t}} t^{-\frac {\nu} {2}}\;+\right.\]
\begin{equation}
\label{ep5.5}
\left.e^{\frac {i\pi(2-\nu)} {2}}
e^{-itx}e^{-\frac {i(k_0^2-k^2)} {4t}} t^{-\frac {\nu} {2}}\right]\;dx\;dt
\end{equation}
We can evaluate the integral in the variable $t$:
\[\int\limits_0^{\infty}
e^{itx}e^{\frac {i\rho} {4t}} t^{-\frac {\nu} {2}}\;dt=
2^{\frac {\nu} {2}} \frac {(x+i0)^{\frac {\nu-2} {4}}} {(\rho+i0)^{\frac {\nu-2} {4}}}
{\cal{K}}_{\frac {\nu-2} {2}}[-i(x+i0)^{1/2}(\rho+i0)^{1/2}]\]
\begin{equation}
\label{ep5.6}
\int\limits_0^{\infty}
e^{-itx}e^{-\frac {i\rho} {4t}} t^{-\frac {\nu} {2}}\;dt=
2^{\frac {\nu} {2}} \frac {(x-i0)^{\frac {\nu-2} {4}}} {(\rho-i0)^{\frac {\nu-2} {4}}}
{\cal{K}}_{\frac {\nu-2} {2}}[i(x-i0)^{1/2}(\rho-i0)^{1/2}]
\end{equation}
where $\rho=k_0^2-k^2$
(Here we have used $\boldsymbol{3.471},(9)$ of ref.\cite{tp8}).
Thus (\ref{ep5.5}) transforms into:
\[f(\rho)=(2\pi)^{\frac {\nu-2} {2}}\int\limits_{-\infty}^{\infty}\hat{f}(x)\left\{
e^{\frac {i\pi(\nu-2)} {4}} \frac {(x+i0)^{\frac {\nu-2} {4}}} {(\rho+i0)^{\frac {\nu-2} {4}}}
{\cal{K}}_{\frac {\nu-2} {2}}[-i(x+i0)^{1/2}(\rho+i0)^{1/2}]\right.+\]
\begin{equation}
\label{ep5.7}
+\left.e^{\frac {i\pi(2-\nu)} {4}} \frac {(x-i0)^{\frac {\nu-2} {4}}} {(\rho-i0)^{\frac {\nu-2} {4}}}
{\cal{K}}_{\frac {\nu-2} {2}}[i(x-i0)^{1/2}(\rho-i0)^{1/2}]\right\}\;dx
\end{equation}
The corresponding inversion formula is then given by:
\[\hat{f}(x)=\frac {1} {(2\pi)^{\frac {\nu+2} {2}}}\int\limits_{-\infty}^{\infty}f(\rho)\left\{
e^{\frac {i\pi(\nu-2)} {4}} \frac {(\rho+i0)^{\frac {\nu-2} {4}}} {(x+i0)^{\frac {\nu-2} {4}}}
{\cal{K}}_{\frac {\nu-2} {2}}[-i(x+i0)^{1/2}(\rho+i0)^{1/2}]\right.+\]
\begin{equation}
\label{ep5.8}
+\left.e^{\frac {i\pi(2-\nu)} {4}} \frac {(\rho-i0)^{\frac {\nu-2} {4}}} {(x-i0)^{\frac {\nu-2} {4}}}
{\cal{K}}_{\frac {\nu-2} {2}}[i(x-i0)^{1/2}(\rho-i0)^{1/2}]\right\}\;d\rho
\end{equation}
Formula (\ref{ep5.7}) is the generalization of Bochner's formula (\ref{ep4.1})
to the Minkowskian Space.

In this case the extension as ultradistribution of $f(\rho)$ to the complex
$\rho$-plane is immediate:
\[F(\rho)=(2\pi)^{\frac {\nu-2} {2}}\int\limits_{-\infty}^{\infty}\hat{f}(x)\left\{
\Theta[\Im(\rho)]
e^{\frac {i\pi(\nu-2)} {4}} \frac {(x+i0)^{\frac {\nu-2} {4}}} {{\rho}^{\frac {\nu-2} {4}}}
{\cal{K}}_{\frac {\nu-2} {2}}[-i(x+i0)^{1/2}{\rho}^{1/2}]\right.-\]
\begin{equation}
\label{ep5.9}
\left.\Theta[-\Im(\rho)]
e^{\frac {i\pi(2-\nu)} {4}} \frac {(x-i0)^{\frac {\nu-2} {4}}} {{\rho}^{\frac {\nu-2} {4}}}
{\cal{K}}_{\frac {\nu-2} {2}}[i(x-i0)^{1/2}{\rho}^{1/2}]\right\}\;dx
\end{equation}
Here we have taken $\rho=\gamma+i\sigma$ and
\begin{equation}
\label{ep5.10}
{\rho}^{1/2}=\sqrt{\frac {\gamma+\sqrt{{\gamma}^2+{\sigma}^2}} {2}}+iSgn(\sigma)
\sqrt{\frac {-\gamma+\sqrt{{\gamma}^2+{\sigma}^2}} {2}}
\end{equation}
Now we can define a Lorentz invariant tempered ultradistribution
as the Fourier transform of a Lorentz invariant distribution of
exponential type. Note that a Lorentz invariant tempered
ultradistribution is not necessarily explicitly Lorentz invariant.
When $\hat{f}$ is a Lorentz invariant distribution of exponential
type, we can use (\ref{ep5.9}) or to adopt the following
treatment: starting from
\begin{equation}
\label{ep5.11}
\iiiint\limits_{-\infty}^{\;\;\;\infty}f(\rho)\phi(\rho,k^0)\;d^4k=
(2\pi)^{\nu}\iiiint\limits_{-\infty}^{\;\;\;\infty}\hat{f}(x)\hat{\phi}(x,x^0)d^4x
\end{equation}
can be deduced the equality:
\[\iint\limits_{-\infty}^{\;\;\;\;\infty}f(\rho)\phi(\rho,k^0)
(k_0^2-\rho)_+^{\frac {\nu-3} {2}}\;d\rho\;dk^0=\]
\begin{equation}
\label{ep5.12}
\iint\limits_{-\infty}^{\;\;\;\;\infty}\hat{f}(x)\hat{\phi}(x,x^0)
(x-x_0^2)_+^{\frac {\nu-3} {2}}\;dx\;dx^0
\end{equation}
Let $g(t)$ defined as:
\begin{equation}
\label{ep5.13}
\hat{g}(t)=\frac {1} {(2\pi)^{\nu}}\int\limits_{-\infty}^{\infty}f(\rho)e^{-i\rho t}\;d\rho
\end{equation}
Then:
\begin{equation}
\label{ep5.14}
F(\rho)=\Theta[\Im(\rho)]\int\limits_0^{\infty}\hat{g}(t)e^{i\rho t}\;dt-
\Theta[-\Im(\rho)]\int\limits_{-\infty}^0\hat{g}(t)e^{i\rho t}\;dt
\end{equation}
or if we use Dirac's formula
\begin{equation}
\label{ep5.15}
F(\rho)=\frac {1} {2\pi i}\int\limits_{-\infty}^{\infty}\frac {f(t)} {t-\rho}\;dt
\end{equation}
The inverse of the Fourier transform can also be evaluated in the following way:
we define:
\[\hat{G}(x,\Lambda)=\frac {1} {(2\pi)^{\frac {\nu+2} {2}}}\oint\limits_{\Gamma}F(\rho)\left\{
e^{\frac {i\pi(\nu-2)} {4}} \frac {(\rho+\Lambda)^{\frac {\nu-2} {4}}} {(x+i0)^{\frac {\nu-2} {4}}}
{\cal{K}}_{\frac {\nu-2} {2}}[-i(x+i0)^{1/2}(\rho+\Lambda)^{1/2}]\right.+\]
\begin{equation}
\label{ep5.16}
+\left.e^{\frac {i\pi(2-\nu)} {4}} \frac {(\rho-\Lambda)^{\frac {\nu-2} {4}}} {(x-i0)^{\frac {\nu-2} {4}}}
{\cal{K}}_{\frac {\nu-2} {2}}[i(x-i0)^{1/2}(\rho-\Lambda)^{1/2}]\right\}\;d\rho
\end{equation}
then
\begin{equation}
\label{ep5.17}
\hat{f}(x)=\hat{G}(x,i0^+)
\end{equation}

\subsection*{Examples}

As a first example we consider the Fourier transform of the function
$e^{a\sqrt{|x_0^2-r^2|}}$ where a is a complex number. The Fourier transform is:
\[F(\rho)=(2\pi)^{\frac {\nu-2} {2}}\int\limits_{-\infty}^{\infty}
e^{|x|^{\frac {1} {2}}}\left\{
\Theta[\Im(\rho)]
e^{\frac {i\pi(\nu-2)} {4}} \frac {(x+i0)^{\frac {\nu-2} {4}}} {{\rho}^{\frac {\nu-2} {4}}}
{\cal{K}}_{\frac {\nu-2} {2}}[-i(x+i0)^{1/2}{\rho}^{1/2}]\right.-\]
\begin{equation}
\label{ep5.18}
\left.\Theta[-\Im(\rho)]
e^{\frac {i\pi(2-\nu)} {4}} \frac {(x-i0)^{\frac {\nu-2} {4}}} {{\rho}^{\frac {\nu-2} {4}}}
{\cal{K}}_{\frac {\nu-2} {2}}[i(x-i0)^{1/2}{\rho}^{1/2}]\right\}\;dx
\end{equation}
Now:
\[e^{\frac {i\pi(\nu-2)}{4}}\int\limits_{-\infty}^{\infty}e^{a|x|^{\frac {1} {2}}}
(x+i0)^{\frac {\nu-2} {4}}{\cal{K}}_{\frac {\nu-2} {2}}[-i(x+i0)^{1/2}{\rho}^{1/2}]=\]
\[2^{\frac {\nu} {2}}\sqrt{\pi}\frac {\Gamma(\nu)} {\Gamma(\frac {\nu+3} {2})}
\frac {e^{\frac {i\pi\nu} {2}}} {({\rho}^{1/2}-ia)^{\nu}}
\boldsymbol{F}\left(\nu,\frac {\nu-1} {2},\frac {\nu+3} {2},
\frac {a-i{\rho}^{1/2}} {a+i{\rho}^{1/2}}\right)-\]
\begin{equation}
\label{ep5.19}
2^{\frac {\nu} {2}}\sqrt{\pi}\frac {\Gamma(\nu)} {\Gamma(\frac {\nu+3} {2})}
\frac {e^{\frac {i\pi\nu} {2}}} {({\rho}^{1/2}+a)^{\nu}}
\boldsymbol{F}\left(\nu,\frac {\nu-1} {2},\frac {\nu+3} {2},
\frac {a+{\rho}^{1/2}} {a-{\rho}^{1/2}}\right)\;\;\Im(\rho)>0
\end{equation}
\[e^{\frac {i\pi(2-\nu)} {4}}\int\limits_{-\infty}^{\infty}e^{a|x|^{\frac {1} {2}}}
(x-i0)^{\frac {\nu-2} {4}}{\cal{K}}_{\frac {\nu-2} {2}}[i(x-i0)^{1/2}{\rho}^{1/2}]=\]
\[2^{\frac {\nu} {2}}\sqrt{\pi}\frac {\Gamma(\nu)} {\Gamma(\frac {\nu+3} {2})}
\frac {e^{-\frac {i\pi\nu} {2}}} {({\rho}^{1/2}+ia)^{\nu}}
\boldsymbol{F}\left(\nu,\frac {\nu-1} {2},\frac {\nu+3} {2},
\frac {a+i{\rho}^{1/2}} {a-i{\rho}^{1/2}}\right)-\]
\begin{equation}
\label{ep5.20}
2^{\frac {\nu} {2}}\sqrt{\pi}\frac {\Gamma(\nu)} {\Gamma(\frac {\nu+3} {2})}
\frac {e^{\frac {i\pi\nu} {2}}} {({\rho}^{1/2}+a)^{\nu}}
\boldsymbol{F}\left(\nu,\frac {\nu-1} {2},\frac {\nu+3} {2},
\frac {a+{\rho}^{1/2}} {a-{\rho}^{1/2}}\right)\;\;\Im(\rho)<0
\end{equation}
To obtain (\ref{ep5.17}) and (\ref{ep5.18}) we have used $\boldsymbol{6.621},(3)$
of ref.\cite{tp8}. With these results we have:
\[F(\rho)=
(4\pi)^{\frac {\nu-1} {2}}\frac {\Gamma(\nu)} {\Gamma(\frac {\nu+3} {2})}
\left\{\Theta[\Im(\rho)]e^{\frac {i\pi\nu} {2}}\left[
\frac {\boldsymbol{F}\left(\nu,\frac {\nu-1} {2},\frac {\nu+3} {2},
\frac {a-i{\rho}^{1/2}} {a+i{\rho}^{1/2}}\right)}
{({\rho}^{1/2}-ia)^{\nu}}\right.\right. -\]
\[\left.\frac {\boldsymbol{F}\left(\nu,\frac {\nu-1} {2},\frac {\nu+3} {2},
\frac {a+{\rho}^{1/2}} {a-{\rho}^{1/2}}\right)}
{({\rho}^{1/2}+a)^{\nu}}\right]-\Theta[-\Im(\rho)]
e^{-\frac {i\pi\nu} {2}}\left[
\frac {\boldsymbol{F}\left(\nu,\frac {\nu-1} {2},\frac {\nu+3} {2},
\frac {a+i{\rho}^{1/2}} {a-i{\rho}^{1/2}}\right)}
{({\rho}^{1/2}+ia)^{\nu}}\right.-\]
\begin{equation}
\label{ep5.21}
\left.\left.
\frac {\boldsymbol{F}\left(\nu,\frac {\nu-1} {2},\frac {\nu+3} {2},
\frac {a+{\rho}^{1/2}} {a-{\rho}^{1/2}}\right)}
{({\rho}^{1/2}+a)^{\nu}}\right]\right\}
\end{equation}
As a second example we evaluate the Fourier transform of the complex mass
Wheeler's propagator.
\begin{equation}
\label{ep5.22}
w_{\mu}(x)=-\frac {i\pi} {2} \frac {{\mu}^{\frac {\nu-2} {2}}}
{(2\pi)^{\frac {\nu} {2}}}x_+^{\frac {2-\nu} {4}} {\cal{J}}_{\frac {2-\nu} {2}}
(\mu x_+^{1/2})
\end{equation}
Then according to (\ref{ep5.9})
\[{\cal{W}}_{\mu}(\rho)=-\frac {i(\mu)^{\frac {\nu-2} {4}}} {4}\int\limits_0^{\infty}
{\cal{J}}_{\frac {2-\nu} {2}}(\mu x^{1/2})\left[
\Theta[\Im(\rho)]
\frac {e^{\frac {i\pi(\nu-2)} {4}}} {{\rho}^{\frac {\nu-2} {4}}}
{\cal{K}}_{\frac {\nu-2} {2}}(-ix^{1/2}{\rho}^{1/2})\right.-\]
\begin{equation}
\label{ep5.23}
\left.\Theta[-\Im(\rho)]
\frac {e^{\frac {i\pi(2-\nu)} {4}}} {{\rho}^{\frac {\nu-2} {4}}}
{\cal{K}}_{\frac {\nu-2} {2}}(ix^{1/2}{\rho}^{1/2})\right]\;dx
\end{equation}
Taking into account that
(See $\boldsymbol{6.576},(3)$, ref.\cite{tp8}):
\[\int\limits_0^{\infty}
{\cal{J}}_{\frac {2-\nu} {2}}(\mu x^{1/2})
{\cal{K}}_{\frac {\nu-2} {2}}(-ix^{1/2}{\rho}^{1/2})\;dx=2{\mu}^{\frac {2-\nu} {2}}
e^{\frac {i\pi(6-\nu)} {4}}\frac {{\rho}^{\frac {\nu-2} {4}}} {\rho-{\mu}^2}\;\Im(\rho)>0\]
\begin{equation}
\label{ep5.24}
\int\limits_0^{\infty}
{\cal{J}}_{\frac {2-\nu} {2}}(\mu x^{1/2})
{\cal{K}}_{\frac {\nu-2} {2}}(ix^{1/2}{\rho}^{1/2})\;dx=2{\mu}^{\frac {2-\nu} {2}}
e^{\frac {i\pi(\nu-6)} {4}}\frac {{\rho}^{\frac {\nu-2} {4}}} {\rho-{\mu}^2}\;\Im(\rho)<0
\end{equation}
we obtain:
\begin{equation}
\label{ep5.25}
{\cal{W}}_{\mu}(\rho)=\frac {i} {2} \frac {Sgn[\Im(\rho)]} {\rho-{\mu}^2}
\end{equation}
As a third example we evaluate the transform of $\delta(x_0^2-r^2)$.
From (\ref{ep5.12}) we obtain:
\begin{equation}
\label{ep5.26}
\iint\limits_{-\infty}^{\;\;\;\;\infty}f(\rho)\phi(\rho,k^0)
(k_0^2-\rho)_+^{\frac {\nu-3} {2}}\;d\rho\;dk^0=
(2\pi)^{\nu}\int\limits_{-\infty}^{\infty}\phi(0,x^0)
|x^0|^{\nu-3}\;dx^0
\end{equation}
According to (\ref{ep5.1}) we can write:
\[\hat{\phi}(x,x^0)=2^{-1}(2\pi)^{-\frac {\nu+1} {2}}(x_0^2-x)_+^{\frac {3-\nu} {4}}
\iint\limits_{-\infty}^{\;\;\;\infty}\phi(\rho,k^0){\cal{J}}_{\frac {\nu-3} {2}}[
(x_0^2-x)_+^{1/2}(k_0^2-\rho)_+^{1/2}]\,\times\]
\begin{equation}
\label{ep5.27}
(k_0^2-\rho)_+^{\frac {\nu-3} {4}}e^{ik_0x^0}\;dk^0\;d\rho
\end{equation}
and consequently:
\[\hat{\phi}(0,x^0)=2^{-1}(2\pi)^{-\frac {\nu+1} {2}}|x^0|^{\frac {3-\nu} {4}}
\iint\limits_{-\infty}^{\;\;\;\infty}\phi(\rho,k^0){\cal{J}}_{\frac {\nu-3} {2}}[
|x^0|^{1/2}(k_0^2-\rho)_+^{1/2}]\,\times\]
\begin{equation}
\label{ep5.28}
(k_0^2-\rho)_+^{\frac {\nu-3} {4}}e^{ik_0x^0}\;dk^0\;d\rho
\end{equation}
Then
\[(2\pi)^{\nu}\int\limits_{-\infty}^{\infty}\phi(0,x^0)
|x^0|^{\nu-3}\;dx^0=2^{-1}(2\pi)^{\frac {\nu-1} {2}}\int\limits_{-\infty}^{\infty}
\phi(\rho,k^0)(k_0^2-\rho)_+^{\frac {\nu-3} {4}}\left[\int\limits_{-\infty}^{\infty}
|x^0|^{\frac {\nu-3} {2}}\right.\times\]
\begin{equation}
\label{ep5.29}
\left.{\cal{J}}_{\frac {\nu-3} {2}}[|x^0|^{1/2}(k_0^2-\rho)_+^{1/2}]
e^{ik_0x^0}\;dx^0\right]\;dk^0\;d\rho
\end{equation}
But
\[\int\limits_{-\infty}^{\infty}
|x^0|^{\frac {\nu-3} {2}}
{\cal{J}}_{\frac {\nu-3} {2}}[|x^0|^{1/2}(k_0^2-\rho)_+^{1/2}]
e^{ik_0x^0}\;dx^0=\]
\begin{equation}
\label{ep5.30}
\frac {2^{\frac {\nu-3} {2}}} {\sqrt{\pi}}\Gamma\left(\frac {\nu-2} {2}\right)
\left[e^{\frac {i\pi(\nu-2)} {2}}
(\rho+i0)^{\frac {2-\nu} {2}}+
e^{\frac {i\pi(2-\nu)} {2}}
(\rho-i0)^{\frac {2-\nu} {2}}\right]
\end{equation}
(See $\boldsymbol{6.623},(1)$,ref\cite{tp8})\\
from which we deduce that
\begin{equation}
\label{ep5.31}
f(\rho)=\frac {(4\pi)^{\frac {\nu-2} {2}}} {2}
\Gamma\left(\frac {\nu-2} {2}\right)
\left[\frac {e^{\frac {i\pi(\nu-2)} {2}}}
{(\rho+i0)^{\frac {\nu-2} {2}}}+
\frac {e^{\frac {i\pi(2-\nu)} {2}}}
{(\rho-i0)^{\frac {\nu-2} {2}}}\right]
\end{equation}
Using then [(\ref{ep5.13}),(\ref{ep5.14})] or (\ref{ep5.15}),
the corresponding ultradistribution is:
\begin{equation}
\label{ep5.32}
F(\rho)=2^{-1}(4\pi)^{\frac {\nu-2} {2}}
\Gamma\left(\frac {\nu-2} {2}\right)
Sgn[\Im(\rho)](-\rho)^{\frac {2-\nu} {2}}
\end{equation}
We proceed now to the calculation of the convolution of two
spherically symmetric tempered ultradistributions.

\section{The Convolution in Euclidean Space}

\setcounter{equation}{0}

The expression for the convolution of two spherically symmetric  functions
was deduced in  ref.\cite{tp12} ($h(k)=(f\ast g)(k)$):
\[h(k)=\frac {2^{4-\nu}{\pi}^{\frac {\nu-1} {2}}}
{\Gamma(\frac {\nu-1} {2})k^{\nu-2}}\iint\limits_{\;0}^{\;\;\;\infty}
f(k_1)g(k_2)\;\times\]
\begin{equation}
\label{ep6.1}
[4k_1^2k_2^2-(k^2-k_1^2-k_2^2)^2]_+^{
\frac {\nu-3} {2}}k_1k_2\;dk_1\;dk_2
\end{equation}
and with the change of variables $\rho=k^2$,${\rho}_1=k_1^2$,
${\rho}_2=k_2^2$ takes the form:
\[h(\rho)=\frac {2^{2-\nu}{\pi}^{\frac {\nu-1} {2}}}
{\Gamma(\frac {\nu-1} {2}){\rho}^{\frac {\nu-2} {2}}}\iint\limits_{\;0}^{\;\;\;\infty}
f({\rho}_1)g({\rho}_2)\;\times\]
\begin{equation}
\label{ep6.2}
[4{\rho}_1{\rho}_2-(\rho-{\rho}_1-{\rho}_2)^2]_+^{
\frac {\nu-3} {2}}\;d{\rho}_1\;d{\rho}_2
\end{equation}
In particular when $\nu=4$ is:
\begin{equation}
\label{ep6.3}
h(\rho)=\frac {\pi}
{2\rho}\iint\limits_{\;0}^{\;\;\;\infty}
f({\rho}_1)g({\rho}_2)
[4{\rho}_1{\rho}_2-(\rho-{\rho}_1-{\rho}_2)^2]_+^{
\frac {1} {2}}\;d{\rho}_1\;d{\rho}_2
\end{equation}
$h(\rho)$ can be extended to complex plane as ultradistribution thus generalizing the
procedure of ref.\cite{tp12}.
According to (\ref{ep4.12}) we can write:
\begin{equation}
\label{ep6.4}
\hspace{-9mm}
\hat{f}(x)\hat{g}(x)=\frac {{\pi}^2} {(2\pi)^6x}
\oint\limits_{{\Gamma}_1}\oint\limits_{{\Gamma}_2}F({\rho}_1)G({\rho}_2){\rho}_1^{1/2}
{\rho}_2^{1/2}{\cal{J}}_1(x^{1/2}{\rho}_1^{1/2}) {\cal{J}}_1(x^{1/2}{\rho}_2^{1/2})
d{\rho}_1\;d{\rho}_2
\end{equation}
and Fourier transforming:
\[
\hspace{-9mm}
{\cal{F}}\left\{\hat{f}(x)\hat{g}(x)\right\}(\rho)=\frac {-{\pi}^2} {(2\pi)^5{\rho}^{1/2}}
\oint\limits_{{\Gamma}_1}\oint\limits_{{\Gamma}_2}F({\rho}_1)G({\rho}_2){\rho}_1^{1/2}
{\rho}_2^{1/2}\left\{\int\limits_0^{\infty}x^{-{1/2}}
{\cal{J}}_1(x^{1/2}{\rho}_1^{1/2})
{\cal{J}}_1(x^{1/2}{\rho}_2^{1/2})\right.\]
\begin{equation}
\label{ep6.5}
\left.\left[
\Theta[\Im(\rho)]{\cal{K}}_1(-ix^{1/2}{\rho}^{1/2})-
\Theta[-\Im(\rho)]{\cal{K}}_1(ix^{1/2}{\rho}^{1/2})\right]dx\right\}
d{\rho}_1\;d{\rho}_2
\end{equation}
The x-integration can be performed with the result:
\[\int\limits_0^{\infty}{\cal{J}}_1(x^{1/2}{\rho}_1^{1/2})
{\cal{J}}_1(x^{1/2}{\rho}_2^{1/2}){\cal{K}}_1(-ix^{1/2}{\rho}^{1/2})dx=\]
\begin{equation}
\label{ep6.6}
-i(\rho{\rho}_1{\rho}_2)^{-1}\left[\rho-{\rho}_1-{\rho}_2-
\sqrt{(\rho-{\rho}_1-{\rho}_2)^2-4{\rho}_1{\rho}_2}\right]\;\;\;\Im(\rho)>0
\end{equation}
\[\int\limits_0^{\infty}{\cal{J}}_1(x^{1/2}{\rho}_1^{1/2})
{\cal{J}}_1(x^{1/2}{\rho}_2^{1/2}){\cal{K}}_1(ix^{1/2}{\rho}^{1/2})dx=\]
\begin{equation}
\label{ep6.7}
i(\rho{\rho}_1{\rho}_2)^{-1}\left[\rho-{\rho}_1-{\rho}_2-
\sqrt{(\rho-{\rho}_1-{\rho}_2)^2-4{\rho}_1{\rho}_2}\right]\;\;\;\Im(\rho)<0
\end{equation}
where we have used $\boldsymbol{6.578},2$ of \cite{tp8} and (7) pag. 238
of \cite{tp11}. Thus
\[H(\rho)=\frac {i\pi} {4\rho}\oint\limits_{{\Gamma}_1}
\oint\limits_{{\Gamma}_2}F({\rho}_1)G({\rho}_2)\;\times\]
\begin{equation}
\label{ep6.8}
\left[\rho-{\rho}_1-{\rho}_2-
\sqrt{(\rho-{\rho}_1-{\rho}_2)^2-4{\rho}_1{\rho}_2}\right]d{\rho}_1\;d{\rho}_2
\end{equation}
$|\Im(\rho)|>|\Im({\rho}_1)|+|\Im({\rho}_2)|$\\
In ref.\cite{tp3} we have defined and shown the existence of the
convolution product between to arbitrary one dimensional tempered
ultradistributions.
Analogously for spherically symmetric ultradistributions we  now define:
\[H_{\lambda}(\rho)=\frac {i\pi} {4\rho}\oint\limits_{{\Gamma}_1}
\oint\limits_{{\Gamma}_2}F({\rho}_1)G({\rho}_2)
{\rho}_1^{\lambda}{\rho}_2^{\lambda}\;\times\]
\begin{equation}
\label{ep6.9}
\left[\rho-{\rho}_1-{\rho}_2-
\sqrt{(\rho-{\rho}_1-{\rho}_2)^2-4{\rho}_1{\rho}_2}\right]d{\rho}_1\;d{\rho}_2
\end{equation}
Let {\textgoth{B}} be a vertical band contained in the complex
${\lambda}$-plane \textgoth{P}.
Integral (\ref{ep6.9}) is an analytic function of $\lambda$ defined in the
domain \textgoth{B}.
Moreover, it is bounded by a power of $|\rho|$.
Then, according to the method of ref.\cite{tp7}, $H_{\lambda}$ can be
analytically continued to other parts of \textgoth{P}.
In particular near the origin we have the Laurent
expansion:
\begin{equation}
\label{ep6.10}
H_{\lambda}(\rho)=\sum\limits_{n=-m}^{\infty}H^{(n)}(\rho){\lambda}^n
\end{equation}
We now define
the convolution product as the $\lambda$-independent term of
(\ref{ep6.10}):
\begin{equation}
\label{ep6.11}
H(\rho)=H^{(0)}(\rho)
\end{equation}
The proof that $H(\rho)$ is a Tempered Ultradistribution is similar
to the one given in ref.\cite{tp3} for the one-dimensional case.
The Fourier antitransform of (\ref{ep6.11}) defines the product of two distributions
of exponential type.
Let ${\hat{H}}_{\lambda}(x)$ be the Fourier antitransform of $H_{\lambda}(\rho)$:
\begin{equation}
\label{ep6.12}
{\hat{H}}_{\lambda}(x)=\sum\limits_{n=-m}^{\infty}{\hat{H}}^{(n)}(x){\lambda}^n
\end{equation}
If we define:
\[{\hat{f}}_{\lambda}(x)={\cal F}^{-1}\{{\rho}^{\lambda}F(\rho)\}\]
\begin{equation}
\label{ep6.13}
{\hat{g}}_{\lambda}(x)={\cal F}^{-1}\{{\rho}^{\lambda}G(\rho)\}
\end{equation}
then
\begin{equation}
\label{ep6.14}
{\hat{H}}_{\lambda}(x)=(2\pi)^4{\hat{f}}_{\lambda}(x){\hat{g}}_{\lambda}(x)
\end{equation}
and taking into account the Laurent developments of $\hat{f}$ and
$\hat{g}$:
\[{\hat{f}}_{\lambda}(x)=\sum\limits_{n=-m_f}^{\infty}{\hat{f}}^{(n)}(x){\lambda}^n\]
\begin{equation}
\label{ep6.15}
{\hat{g}}_{\lambda}(x)=\sum\limits_{n=-m_g}^{\infty}{\hat{g}}^{(n)}(x){\lambda}^n
\end{equation}
we can write:
\begin{equation}
\label{ep6.16}
\sum\limits_{n=-m}^{\infty}{\hat{H}}^{(n)}(x){\lambda}^n=(2\pi)^4
\sum\limits_{n=-m}^{\infty}\left(\sum\limits_{k=-m_f}^{n+m_g} {\hat{f}}^{(k)}(x)
{\hat{g}}^{(n-k)}(x)\right){\lambda}^n
\end{equation}
$(m=m_f+m_g)$\\
and as a consequence:
\begin{equation}
\label{ep6.17}
{\hat{H}}^{(0)}(x)=
\sum\limits_{k=-m_f}^{m_g}{\hat{f}}^{(k)}(x){\hat{g}}^{(-k)}(x)
\end{equation}
We will give now some examples of the use of (\ref{ep6.11}) and (\ref{ep6.17}).

\subsection*{Examples}

As a first example we evaluate the convolution of two Dirac's delta of complex mass:
\[\left(\delta(\rho-{\mu}^2)=-\frac {1} {2\pi i(\rho-{\mu}^2)}\right)\]
According to (\ref{ep6.9}),(\ref{ep6.10}),(\ref{ep6.11}) we have:
\[\delta(\rho-{\mu}_1^2)\ast\delta(\rho-{\mu}_2^2)=\frac {i\pi} {4\rho}
\left[\rho-{\mu}_1^2-{\mu}_2^2-\sqrt{(\rho-{\mu}_1^2-{\mu}_2^2)^2-4{\mu}_1^2
{\mu}_2^2}\right]\]
As an ultradistribution only the term containing the square root is different from
zero (cf.(\ref{ep4.11})). We then have:
\begin{equation}
\label{ep6.18}
\delta(\rho-{\mu}_1^2)\ast\delta(\rho-{\mu}_2^2)=
-\frac {i\pi} {4{\rho}}
\sqrt{(\rho-{\mu}_1^2-{\mu}_2^2)^2-4{\mu}_1^2
{\mu}_2^2}
\end{equation}
When ${\mu}_1={\mu}_2=m$ ($m$ real) we obtain:
\begin{equation}
\label{ep6.19}
\delta(\rho-m^2)\ast\delta(\rho-m^2)=-\frac {i\pi} {4{\rho}^{1/2}}
\sqrt{\rho-4m^2}
\end{equation}
As a second example we evaluate the convolution of two massless
Feynman's propagators. We have:
\[f(\rho)=\frac {1} {\rho}\]
\[F(\rho)=-\frac {1} {2\pi i\rho}\ln(-\rho)\]
\[F_{\lambda}(\rho)=-\frac {1} {2\pi i} {\rho}^{\lambda-1}\ln(-\rho)\]
\[{\hat{f}}_{\lambda}(x)=\frac {1} {8{\pi}^2x^{1/2}}
\oint\limits_{\Gamma}\left(-\frac {1} {2\pi i}{\rho}^{\lambda-1}
\ln(-\rho)\right){\rho}^{1/2}{\cal J}_1(x^{1/2}{\rho}^{1/2})\;d\rho=\]
\[\frac {2^{2\lambda}\Gamma(1+\lambda)}
{4{\pi}^2\Gamma(1-\lambda)}x^{-\lambda-1}-e^{i\pi\lambda}\sin(\pi\lambda)
\frac {2^{2\lambda}\Gamma(1+\lambda)}
{4{\pi}^2\Gamma(1-\lambda)}x^{-\lambda-1}\left[i\pi+\right.\]
\begin{equation}
\label{ep6.20}
\left. 2\ln(2)+\psi(1+\lambda)+\psi(1-\lambda)-\ln(x)\right]
\end{equation}
where $\psi(z)={\Gamma}^{'}(z)/\Gamma(z)$\\
From (\ref{ep6.20}) we have
\begin{equation}
\label{ep6.21}
{\hat{f}}_{\lambda}(x)=(2\pi)^{-2} x^{-1}+S_{\lambda}(x)
\end{equation}
with
\[\lim_{\lambda\rightarrow 0}S_{\lambda}(x)=0\]
Then
\begin{equation}
\label{ep6.22}
{\hat{f}}_{\lambda}^2(x)=(2\pi)^{-4} x^{-2}+T_{\lambda}(x)
\end{equation}
with
\[\lim_{\lambda\rightarrow 0}T_{\lambda}(x)=0\]
As a consequence
\begin{equation}
\label{ep6.23}
{\hat{f}}^2(x)=(2\pi)^{-4} x^{-2}
\end{equation}
Taking into account that
\[{\cal F}\{x^{-2}\}=-{\pi}^2\ln(\rho)\]
we obtain
\begin{equation}
\label{ep6.24}
\frac {1} {\rho}\ast\frac {1} {\rho}=-{\pi}^2\ln(\rho)
\end{equation}

\section{The Convolution in Minkowskian space}

In this section we deduce the formula for the convolution of two Lorentz invariant
functions and then we consider the central topic of this paper, i.e: the convolution of two
Lorentz invariant tempered ultradistributions.

\setcounter{equation}{0}

\subsection{The generalization of Dimensional Regularization in Configuration Space
to the Minkowskian Space}

The convolution of two Lorentz invariant functions is given by:
\begin{equation}
\label{ep7.1}
\{f\ast g\}(p_{\mu}^2)=\int\limits_{-\infty}^{\infty}\cdot\cdot\cdot
\int\limits_{-\infty}^{\infty}
f[(p_{\mu}-k_{\mu})^2]
g(k_{\mu}^2)\;d^{\nu}k
\end{equation}
and can be re-written as
\begin{equation}
\label{ep7.2}
\int\limits_{-\infty}^{\infty}\cdot\cdot\cdot
\int\limits_{-\infty}^{\infty}f({\eta}_1)
g({\eta}_2)\delta[{\eta}_1-(p_{\mu}-k_{\mu})^2]\delta({\eta}_2-k_{\mu}^2)
\;d{\eta}_1\;d{\eta}_2\;d^{\nu}k
\end{equation}
We select the axis of coordinates in a way that the spatial component
of $p_{\mu}$, $\vec{p}$ coincides with the first spatial coordinate
($p_{\mu}^2=p_0^2-p_1^2$). Then we have:
\begin{equation}
\label{ep7.3}
\hspace{-5mm}
\frac {{\pi}^{\frac {\nu-2} {2}}} {2|p_0|}
\iiint\limits_{-\infty}^{\;\;\;\infty}\frac {f({\eta}_1)g({\eta}_2)}
{\Gamma\left(\frac {\nu-2} {2}\right)}
\left[\frac {(p_{\mu}^2-{\eta}_1+{\eta}_2+2p_1k_1)^2} {4p_0^2}-k_1^2-{\eta}_2
\right]^{\frac {\nu-4} {2}}\!\!\!\!\!\!
d{\eta}_1\;d{\eta}_2\;dk_1
\end{equation}
Using:
\begin{equation}
\label{ep7.4}
x_+^{\frac {\nu-4} {2}}=\frac {\Gamma\left(\frac {\nu-2} {2}\right)
e^{i\pi(\frac {2-\nu} {4})}} {2\pi}
\int\limits_{-\infty}^{\infty}(t-i0)^{\frac {3-\nu} {2}}
e^{itx}\;dt
\end{equation}
with
\begin{equation}
\label{ep7.5}
x=-4p_{\mu}^2k_1^2+4p_1k_1(p_{\mu}^2-{\eta}_1+{\eta}_2 )+
(p_{\mu}^2-{\eta}_1+{\eta}_2)^2-4p_0^2{\eta}_2
\end{equation}
we can evaluate the integral in the variable $k_1$ using
$\boldsymbol{2.462} (1)$ of ref.\cite{tp8}. The result is:
\begin{equation}
\label{ep7.6}
\sqrt{2\pi}[i(8tp_{\mu}^2-i0)]^{-\frac {1} {2}}
e^{\frac {itp_1^2(p_{\mu}^2-{\eta}_1+{\eta}_2)} {p_{\mu}^2}}
\end{equation}
We can now perform the $t$ integration:
\begin{equation}
\label{ep7.7}
\hspace{-5mm}
I=\lim_{\epsilon\rightarrow 0}\frac {\Gamma\left(\frac {\nu-2} {2}\right)
e^{\frac {i\pi(1-\nu)} {4}}} {4\sqrt{\pi}}\int\limits_{-\infty}^{\infty}
(t-i\epsilon)^{\frac {2-\nu} {2}}(tp_{\mu}^2-i\epsilon)^{-\frac {1} {2}}
e^{\frac {itp_0^2[(p_{\mu}^2-{\eta}_1+{\eta}_2)^2-4p_{\mu}^2{\eta}_2]}
{p_{\mu}^2}}dt
\end{equation}
Formula (\ref{ep7.7}) is defined for $\nu=2n$. In this case (\ref{ep7.7})
is proportional to the derivative of the same order of the Dirac's formula for
\[(tp_{\mu}^2-i0)^{-\frac {1} {2}}
e^{\frac {itp_0^2[(p_{\mu}^2-{\eta}_1+{\eta}_2)^2-4p_{\mu}^2{\eta}_2]}
{p_{\mu}^2}}\]
with $z=i\epsilon$. Thus we have
\[I=\frac {\Gamma\left(\frac {\nu-2} {2}\right)
e^{\frac {i\pi(1-\nu)} {4}}} {4\sqrt{\pi}}\int\limits_{-\infty}^{\infty}
(p_{\mu}^2-i0)^{-\frac {1} {2}} t_+^{\frac {1-\nu} {2}}
e^{\frac {itp_0^2[(p_{\mu}^2-{\eta}_1+{\eta}_2)^2-4p_{\mu}^2{\eta}_2]}
{p_{\mu}^2}}+\]
\begin{equation}
\label{ep7.8}
(p_{\mu}^2+i0)^{-\frac {1} {2}} t_+^{\frac {1-\nu} {2}}
e^{-\frac {itp_0^2[(p_{\mu}^2-{\eta}_1+{\eta}_2)^2-4p_{\mu}^2{\eta}_2]}
{p_{\mu}^2}}dt
\end{equation}
The result of (\ref{ep7.8})is immediate (is a Fourier transform).
We consider first the case $\nu\neq 2n+1$:
\[I=\frac {e^{\frac {i\pi(2-\nu)} {2}}} {4\sqrt{\pi}}
\Gamma\left(\frac {\nu-2} {2}\right)
\Gamma\left(\frac {3-\nu} {2}\right)
|p_0|^{\nu-3}\;\;\;\times\]
\[\left\{(p_{\mu}^2-i0)^{-\frac {1} {2}}
\left[\frac {(p_{\mu}^2-{\eta}_1+{\eta}_2)^2-4p_{\mu}^2{\eta}_2}
{p_{\mu}^2}+i0\right]^{\frac {\nu-3} {2}}\right.\]
\begin{equation}
\label{ep7.9}
+e^{i\pi(\nu-2)}
\left.(p_{\mu}^2+i0)^{-\frac {1} {2}}
\left[\frac {(p_{\mu}^2-{\eta}_1+{\eta}_2)^2-4p_{\mu}^2{\eta}_2}
{p_{\mu}^2}-i0\right]^{\frac {\nu-3} {2}}\right\}
\end{equation}
With this result we have for (\ref{ep7.3})
\[h(\rho)=
\frac {{\pi}^{\frac {\nu-3} {2}}} {2^{\nu-1}}
e^{\frac {i\pi(2-\nu)} {2}}
\Gamma\left(\frac {3-\nu} {2}\right)
\iint\limits_{-\infty}^{\;\;\;\infty}f({\rho}_1)g({\rho}_2)
\;\;\;\times\]
\[\left\{(\rho-i0)^{-\frac {1} {2}}
\left[\frac {(\rho-{\rho}_1-{\rho}_2)^2-4{\rho}_1{\rho}_2}
{\rho}+i0\right]^{\frac {\nu-3} {2}}+e^{i\pi(\nu-2)}\right.\;\;\;\times\]
\begin{equation}
\label{ep7.10}
\left.(\rho+i0)^{-\frac {1} {2}}
\left[\frac {(\rho-{\rho}_1-{\rho}_2)^2-4{\rho}_1{\rho}_2}
{\rho}-i0\right]^{\frac {\nu-3} {2}}\right\}d{\rho}_1\;d{\rho}_2
\end{equation}
where $\rho=p_{\mu}^2$ and $h=f\ast g$.\\
When $\nu=4$ we have
\begin{equation}
\label{ep7.11}
h(\rho)=\frac {\pi} {2\rho}\iint\limits_{-\infty}^{\;\;\;\infty}
f({\rho}_1)g({\rho}_2)
\left[(\rho-{\rho}_1-{\rho}_2)^2-4{\rho}_1{\rho}_2\right]_+^{\frac {1} {2}}
\;d{\rho}_1\;d{\rho}_2
\end{equation}
When $\nu=2n+1$ we obtain:
\[h(\rho)=-\frac {i{\pi}^{n-1}} {2^{2n}(n-1)!}
\iint\limits_{-\infty}^{\;\;\;\infty}f({\rho}_1)g({\rho}_2)\left[\frac{(\rho-{\rho}_1-
{\rho}_2)^2-4{\rho}_1{\rho}_2} {\rho}\right]^{n-1}\left\{(\rho-i0)^{-\frac {1} {2}}
\right.\times\]
\[\left[\psi(n)+\frac {i\pi} {2}+\ln
\left[\frac{(\rho-{\rho}_1-
{\rho}_2)^2-4{\rho}_1{\rho}_2} {\rho}+i0\right]\right]-(\rho+i0)^{-\frac {1} {2}}\]
\begin{equation}
\label{ep7.12}
\left.\left[\psi(n)+\frac {i\pi} {2}+\ln
\left[-\frac{(\rho-{\rho}_1-
{\rho}_2)^2-4{\rho}_1{\rho}_2} {\rho}+i0\right]\right]\right\}d{\rho}_1\;d{\rho}_2
\end{equation}
As an example we will evaluate the convolution of $\delta(\rho-m_1^2)$
with $\delta(\rho-m_2^2)$ for $\nu\neq 2n+1$. In this case we have:
\[h(\rho)=
\frac {{\pi}^{\frac {\nu-3} {2}}} {2^{\nu-1}}
e^{\frac {i\pi(2-\nu)} {2}}
\left\{(\rho-i0)^{-\frac {1} {2}}
\left[\frac {(\rho-m_1^2-m_2^2)^2-4m_1^2m_2^2}
{\rho}+i0\right]^{\frac {\nu-3} {2}}\right.+\]
\begin{equation}
\label{ep7.13}
\left.e^{i\pi(\nu-2)}(\rho+i0)^{-\frac {1} {2}}
\left[\frac {(\rho-m_1^2-m_2^2)^2-4m_1^2m_2^2}
{\rho}-i0\right]^{\frac {\nu-3} {2}}\right\}
\end{equation}
When $\nu=4, m_1=0, m_2=m$ we obtain:
\begin{equation}
\label{ep7.14}
\delta(\rho)\ast\delta(\rho-m^2)=\frac {\pi} {2\rho} |\rho-m^2|
\end{equation}
If we use the dimension $\nu$ as a regularizing parameter, we can define
the product of two tempered distributions as
\[\hat{h}(x,\nu)=(2\pi)^{\nu}\hat{f}(x,\nu)\hat{g}(x,\nu)=
(2\pi)^{\nu}{\cal F}^{-1}\{f(\rho,\nu)\}{\cal F}^{-1}\{g(\rho,\nu)\}=\]
\begin{equation}
\label{ep7.15}
{\cal F}^{-1}\{f(\rho,\nu)\ast g(\rho,\nu)\}={\cal F}^{-1}\{h(\rho,\nu)\}
\end{equation}
where ${\cal F}^{-1}$ was defined in section 5 by means of (\ref{ep5.8})
and where (\ref{ep7.10}) should be re-interpreted
as:
\[h(\rho,\nu)=
\frac {{\pi}^{\frac {\nu-3} {2}}} {2^{\nu-1}}
e^{\frac {i\pi(2-\nu)} {2}}
\Gamma\left(\frac {3-\nu} {2}\right)
\iint\limits_{-\infty}^{\;\;\;\infty}f({\rho}_1,\nu)g({\rho}_2,\nu)
\;\;\;\times\]
\[\left\{(\rho-i0)^{-\frac {1} {2}}
\left[\frac {(\rho-{\rho}_1-{\rho}_2)^2-4{\rho}_1{\rho}_2}
{\rho}+i0\right]^{\frac {\nu-3} {2}}+e^{i\pi(\nu-2)}\right.\;\;\;\times\]
\begin{equation}
\label{ep7.16}
\left.(\rho+i0)^{-\frac {1} {2}}
\left[\frac {(\rho-{\rho}_1-{\rho}_2)^2-4{\rho}_1{\rho}_2}
{\rho}-i0\right]^{\frac {\nu-3} {2}}\right\}d{\rho}_1\;d{\rho}_2
\end{equation}
The same procedure is valid when $\hat{f}(x,\nu)$ and $\hat{g}(x,\nu)$ are
distributions of exponential type. Here $f(\rho,\nu)$ and $g(\rho,\nu)$ are
defined by:
\[F(\rho,\nu)=\frac {1} {2\pi i}\int\limits_{-\infty}^{\infty}
\frac {f(t,\nu)} {t-\rho} dt\]
\[G(\rho,\nu)=\frac {1} {2\pi i}\int\limits_{-\infty}^{\infty}
\frac {g(t,\nu)} {t-\rho} dt\]
where $F$ and $G$ are the tempered ultradistributions given by
\[F(\rho,\nu)={\cal F}\{\hat{f}(x,\nu)\}\;\;\;\;\;\;\;
G(\rho,\nu)={\cal F}\{\hat{g}(x,\nu)\}\]
This procedure generalize to the Minkowskian space the dimensional
regularization in configuration space defined in ref.\cite{tp12} for
the Euclidean space. As an example of the use of this method
we give the evaluation of the convolution product of two complex mass
Wheeler's propagators.  From (\ref{ep5.22}) and (\ref{ep5.9}) we have:
\[{\cal F}\left\{w_{{\mu}_1}(x,\nu)w_{{\mu}_2}(x,\nu)\right\}=
-\frac {{\pi}^2} {2\rho}\frac {({\mu}_1{\mu}_2)^{\frac {\nu-2} {2}}}
{(2\pi)^{\frac {\nu+2} {2}}}\int\limits_0^{\infty}x^{\frac {4-\nu} {2}}
{\cal J}_{\frac {2-\nu} {2}}({\mu}_1x){\cal J}_{\frac {2-\nu} {2}}({\mu}_2x)\;\;\;\times\]
\begin{equation}
\label{ep7.17}
\hspace{-1mm}
\left\{\Theta[\Im(\rho)]e^{\frac {i\pi(\nu-2)} {4}}{\cal K}_{\frac {\nu-2} {2}}(-ix{\rho}^{1/2})-
\Theta[-\Im(\rho)]e^{\frac {i\pi(2-\nu)} {4}}{\cal K}_{\frac {\nu-2} {2}}(ix{\rho}^{1/2})
\right\}dx
\end{equation}
To evaluate (\ref{ep7.17}) we use
\[\int\limits_0^{\infty}
{\cal J}_{\frac {2-\nu} {2}}({\mu}_1x){\cal J}_{\frac {2-\nu} {2}}({\mu}_2x)
{\cal K}_{\frac {\nu-2} {2}}(xz)\;dx=\]
\begin{equation}
\label{ep7.18}
\frac {1} {\sqrt{\pi}}
\frac {\Gamma\left(\frac {3-\nu} {2}\right)} {2^{\frac {3\nu-6} {2}}}
\frac {z^{\frac {2-\nu} {2}}} {({\mu}_1{\mu}_2)^{\frac {\nu-2} {2}}}
\left[(z^2+{\mu}_1^2+{\mu}_2^2)^2-4{\mu}_1^2{\mu}_2^2\right]^{\frac {
\nu-3} {2}}
\end{equation}
and to deduce (\ref{ep7.18}) we have used:
\[{\cal K}_{\frac {\nu-2} {2}}(xz)=\frac {1} {2}
\left(\frac {zx} {2}\right)^{\frac {\nu-2} {2}}
\int\limits_0^{\infty}t^{-\frac {\nu} {2}}
e^{-t-{\frac {z^2x^2} {4t}}}dt\]
(See $\boldsymbol{8.432} (6)$ of ref.\cite{tp8}). Thus from (\ref{ep7.18})
we have:
\[{\cal F}\left\{w_{{\mu}_1}(x,\nu)w_{{\mu}_2}(x,\nu)\right\}=
\frac {(2\pi)^{\frac {1-\nu} {2}}} {2^{\frac {3\nu-1} {2}}}
\Gamma\left(\frac {3-\nu} {2}\right)
e^{\frac {i\pi(\nu-2)} {2}}\;\;\;\times\]
\begin{equation}
\label{ep7.19}
{\rho}^{\frac {\nu-2} {2}} Sgn[\Im(\rho)]\left[
(\rho-{\mu}_1^2-{\mu}_2^2)^2-4{\mu}_1^2{\mu}_2^2\right]^{\frac {
\nu-3} {2}}
\end{equation}
and consequently:
\[\left\{W_{{\mu}_1}(\rho,\nu)\ast W_{{\mu}_2}(\rho,\nu)\right\}=
\frac {(2\pi)^{\frac {\nu+1} {2}}} {2^{\frac {3\nu-1} {2}}}
\Gamma\left(\frac {3-\nu} {2}\right)
e^{\frac {i\pi(\nu-2)} {2}}\;\;\;\times\]
\begin{equation}
\label{ep7.20}
{\rho}^{\frac {\nu-2} {2}} Sgn[\Im(\rho)]\left[
(\rho-{\mu}_1^2-{\mu}_2^2)^2-4{\mu}_1^2{\mu}_2^2\right]^{\frac {
\nu-3} {2}}
\end{equation}

\subsection{The Convolution of two Lorentz Invariant Tempered Ultradistributions}

To obtain a expression for the convolution of two tempered ultradistributions
we consider the formula (\ref{ep7.11}). As a first step we extend $h(\rho)$
as tempered ultradistribution. For this pourpose we consider the function:
\begin{equation}
\label{ep7.21}
l(\rho,{\rho}_1,{\rho}_2)=
\left[(\rho-{\rho}_1-{\rho}_2)^2-4{\rho}_1{\rho}_2\right]_+^{\frac {1} {2}}
\end{equation}
The Fourier antitransform of (\ref{ep7.21}) is:
\[\hat{l}(x,{\rho}_1,{\rho}_2)=
\frac {e^{-i({\rho}_1+{\rho}_2)x}} {|x|}\left\{({\rho}_1{\rho}_2+i0)^{
\frac {1} {2}}{\cal N}_1\left[2({\rho}_1{\rho}_2+i0)^{\frac {1} {2}}|x|\right]+
\right.\]
\begin{equation}
\label{ep7.22}
\left.\Theta(-{\rho}_1{\rho}_2)\sqrt{-{\rho}_1{\rho}_2}
{\cal J}_1(2i\sqrt{-{\rho}_1{\rho}_2}\;|x|)\right\}
\end{equation}
where ${\cal N}_1$ is the Newman function.
If we consider now the distribution:
\begin{equation}
\label{ep7.23}
m(\rho,{\rho}_1,{\rho}_2)={\rho}^{-1}
\left[(\rho-{\rho}_1-{\rho}_2)^2-4{\rho}_1{\rho}_2\right]_+^{\frac {1} {2}}
\end{equation}
the corresponding tempered ultradistribution is:
\begin{equation}
\label{ep7.24}
M(\rho,{\rho}_1,{\rho}_2)=\frac {1} {2\pi i}\int\limits_{-\infty}^{\infty}
\frac {t^{-1}
\left[(t-{\rho}_1-{\rho}_2)^2-4{\rho}_1{\rho}_2\right]_+^{\frac {1} {2}}}
{t-\rho}\;dt
\end{equation}
which can also be written as:
\[M(\rho,{\rho}_1,{\rho}_2)=
\frac {1} {\rho}\left\{{\cal F}\{\hat{l}\}(\rho,{\rho}_1,{\rho}_2)\right.\;\;-\]
\begin{equation}
\label{ep7.25}
\left.\frac {1} {2}\left[{\cal F}\{\hat{l}\}(i0,{\rho}_1,{\rho}_2)+
{\cal F}\{\hat{l}\}(-i0,{\rho}_1,{\rho}_2)
\right]\right\}
\end{equation}
Thus the extension to the complex plane of $h(\rho)$, $N(\rho)$ is:
\begin{equation}
\label{ep7.26}
N(\rho)=\frac {\pi} {2}\iint\limits_{-\infty}^{\;\;\;\infty}
f({\rho}_1)g({\rho}_2)
M(\rho,{\rho}_1,{\rho}_2)\;d{\rho}_1\;d{\rho}_2
\end{equation}
To obtain $M$ in an explicit way we use the following Laplace
transforms:
\[{\cal L}\left\{t^{-1}{\cal N}_1(at)\right\}(s)=
-\frac {2} {\pi a}\sqrt{s^2+a^2}\ln\left(\frac {\sqrt{s^2+a^2}+s} {a}\right)\;\;+\]
\begin{equation}
\label{ep7.27}
\frac {2s} {a\pi}\left(\ln(2)+1-\gamma\right)
\end{equation}
\begin{equation}
\label{ep7.28}
{\cal L}\left\{t^{-1}{\cal J}_1(at)\right\}(s)=
\frac {\sqrt{s^2+a^2}-s} {a}
\end{equation}
(see \cite{tp13} pags. 310 1nd 313). Then we have for the Fourier transforms:
\[{\cal F}\left\{|t|^{-1}{\cal N}_1(a|t|)\right\}(\rho)=
-\frac {2} {\pi a}\left\{\Theta[\Im(\rho)]\left[
\sqrt{a^2-{\rho}^2}\ln\left(\frac {\sqrt{a^2-{\rho}^2}-i\rho} {a}\right)\right.\right.+\]
\[\left.i\rho\left(\ln(2)+1-\gamma\right)\right]-\Theta[-\Im(\rho)]\left[
\sqrt{a^2-{\rho}^2}\ln\left(\frac {\sqrt{a^2-{\rho}^2}+i\rho} {a}\right)\right.-\]
\begin{equation}
\label{ep7.29}
\left.\left.i\rho\left(\ln(2)+1-\gamma\right)\right]\right\}
\end{equation}
\[{\cal F}\left\{|t|^{-1}{\cal J}_1(a|t|)\right\}(\rho)=\Theta[\Im(\rho)]
\frac {\sqrt{a^2-{\rho}^2}-i\rho} {a}\;\;-\]
\begin{equation}
\label{ep7.30}
\Theta[-\Im(\rho)]
\frac {\sqrt{a^2-{\rho}^2}+i\rho} {a}
\end{equation}
With these results we obtain:
\[M(\rho)=\Theta[\Im(\rho)]\left\{\Theta({\rho}_1{\rho}_2)
\sqrt{4{\rho}_1{\rho}_2-(\rho-{\rho}_1-{\rho}_2)^2}\;\;\times\right.\]
\[\ln\left[\frac{\sqrt{4{\rho}_1{\rho}_2-(\rho-{\rho}_1-{\rho}_2)^2}
-i(\rho-{\rho}_1-{\rho}_2)} {2\sqrt{{\rho}_1{\rho}_2}}\right]\;\;
+ \]
\[\Theta(-{\rho}_1{\rho}_2)\left\{\frac {i\pi} {2}\left[
\sqrt{4{\rho}_1{\rho}_2-(\rho-{\rho}_1-{\rho}_2)^2}-i(\rho-{\rho}_1-{\rho}_2)
\right]\right.\;\;+\]
\[\left.\left. \sqrt{4{\rho}_1{\rho}_2-(\rho-{\rho}_1-{\rho}_2)^2}
\ln\left[\frac{\sqrt{4{\rho}_1{\rho}_2-(\rho-{\rho}_1-{\rho}_2)^2}
-i(\rho-{\rho}_1-{\rho}_2)} {2i\sqrt{-{\rho}_1{\rho}_2}}\right]\right\}\right\}-\]
\[\Theta[-\Im(\rho)]\left\{\Theta({\rho}_1{\rho}_2)
\sqrt{4{\rho}_1{\rho}_2-(\rho-{\rho}_1-{\rho}_2)^2}\right.\times\]
\[\ln\left[\frac{\sqrt{4{\rho}_1{\rho}_2-(\rho-{\rho}_1-{\rho}_2)^2}
+i(\rho-{\rho}_1-{\rho}_2)} {2\sqrt{{\rho}_1{\rho}_2}}\right]\;\;
+ \]
\[\Theta(-{\rho}_1{\rho}_2)\left\{\frac {i\pi} {2}\left[
\sqrt{4{\rho}_1{\rho}_2-(\rho-{\rho}_1-{\rho}_2)^2}+i(\rho-{\rho}_1-{\rho}_2)
\right]\right.\;\;+\]
\[\left.\left.\sqrt{4{\rho}_1{\rho}_2-(\rho-{\rho}_1-{\rho}_2)^2}
\ln\left[\frac{\sqrt{4{\rho}_1{\rho}_2-(\rho-{\rho}_1-{\rho}_2)^2}
+i(\rho-{\rho}_1-{\rho}_2)} {2i\sqrt{-{\rho}_1{\rho}_2}}\right]\right\}\right\}-\]
\[\frac {i} {2}\left\{\Theta({\rho}_1{\rho}_2)({\rho}_1-{\rho}_2)
\ln\left(\frac {{\rho}_1} {{\rho}_2}\right)+
\Theta(-{\rho}_1{\rho}_2)({\rho}_1-{\rho}_2)
\ln\left(-\frac {{\rho}_1} {{\rho}_2}\right)\right.+\]
\[\Theta(-{\rho}_1)\Theta({\rho}_2)\left[i\pi({\rho}_1-{\rho}_2)
Sgn({\rho}_1+{\rho}_2)+
2i\pi{\rho}_2\Theta({\rho}_1+{\rho}_2)+
2i\pi{\rho}_1\Theta(-{\rho}_1-{\rho}_2)\right]+\]
\[\Theta({\rho}_1)\Theta(-{\rho}_2)\left[-i\pi({\rho}_1-{\rho}_2)
Sgn({\rho}_1+{\rho}_2)+
2i\pi{\rho}_1\Theta({\rho}_1+{\rho}_2)\right.\;\;+\]
\begin{equation}
\label{ep7.31}
\left.\left.2i\pi{\rho}_2\Theta(-{\rho}_1-{\rho}_2)\right]
\right\}
\end{equation}
To obtain an expression for the convolution of two ultradistribution
we use for the Heaviside function the identity:
\begin{equation}
\label{ep7.32}
\Theta(xy)=\Theta(x)\Theta(y)+\Theta(-x)\Theta(-y)
\end{equation}
Taking into account that
\begin{equation}
\label{ep7.33}
\Theta(\rho)=\lim_{\Lambda\rightarrow i0^+}\frac {1} {2\pi i}
[\ln(-\rho+\Lambda)-\ln(-\rho-\Lambda)]
\end{equation}
a conceptually simple by rather lengthy expression is obtained
for Lorentz invariant tempered ultradistributions:
\[H_{\lambda}(\rho, \Lambda)=\frac {1} {8{\pi}^2\rho}
\int\limits_{{\Gamma}_1}\int\limits_{{\Gamma}_2}F({\rho}_1)G({\rho}_2)
{\rho}_1^{\lambda}{\rho}_2^{\lambda}\left\{\Theta[\Im(\rho)]\left\{[
\ln(-{\rho}_1+\Lambda)-\ln(-{\rho}_1-\Lambda)]\times
\right.\right.\]
\[[\ln(-{\rho}_2+\Lambda)-\ln(-{\rho}_2-\Lambda)]
\sqrt{4({\rho}_1+\Lambda)({\rho}_2+\Lambda)-
(\rho-{\rho}_1-{\rho}_2-2\Lambda)^2}\times\]
\[\ln\left[\frac {\sqrt{4({\rho}_1+\Lambda)({\rho}_2+\Lambda)-
(\rho-{\rho}_1-{\rho}_2-2\Lambda)^2}-i(\rho-{\rho}_1-{\rho}_2-2\Lambda)}
{2\sqrt{({\rho}_1+\Lambda)({\rho}_2+\Lambda)}}\right]+ \]
\[[\ln({\rho}_1+\Lambda)-\ln({\rho}_1-\Lambda)]
[\ln({\rho}_2+\Lambda)-\ln({\rho}_2-\Lambda)]\times\]
\[\sqrt{4({\rho}_1-\Lambda)({\rho}_2-\Lambda)-
(\rho-{\rho}_1-{\rho}_2+2\Lambda)^2}\times\]
\[\ln\left[\frac {\sqrt{4({\rho}_1-\Lambda)({\rho}_2-\Lambda)-
(\rho-{\rho}_1-{\rho}_2+2\Lambda)^2}-i(\rho-{\rho}_1-{\rho}_2+2\Lambda)}
{2\sqrt{({\rho}_1-\Lambda)({\rho}_2-\Lambda)}}\right]+ \]
\[[\ln({\rho}_1+\Lambda)-\ln({\rho}_1-\Lambda)]
[\ln(-{\rho}_2+\Lambda)-\ln(-{\rho}_2-\Lambda)]\times\]
\[\left\{\frac {i\pi} {2}\left[
\sqrt{4({\rho}_1+\Lambda)({\rho}_2-\Lambda)-
(\rho-{\rho}_1-{\rho}_2)^2}-i(\rho-{\rho}_1-{\rho}_2)\right]\right.+\]
\[\sqrt{4({\rho}_1+\Lambda)({\rho}_2-\Lambda)-
(\rho-{\rho}_1-{\rho}_2)^2}\times\]
\[\left.\ln\left[\frac {\sqrt{4({\rho}_1+\Lambda)({\rho}_2-\Lambda)-
(\rho-{\rho}_1-{\rho}_2)^2}-i(\rho-{\rho}_1-{\rho}_2)}
{2i\sqrt{-({\rho}_1+\Lambda)({\rho}_2-\Lambda)}}\right]\right\}+ \]
\[[\ln(-{\rho}_1+\Lambda)-\ln(-{\rho}_1-\Lambda)]
[\ln({\rho}_2+\Lambda)-\ln({\rho}_2-\Lambda)]\times\]
\[\left\{\frac {i\pi} {2}\left[
\sqrt{4({\rho}_1-\Lambda)({\rho}_2+\Lambda)-
(\rho-{\rho}_1-{\rho}_2)^2}-i(\rho-{\rho}_1-{\rho}_2)\right]\right.+\]
\[\sqrt{4({\rho}_1-\Lambda)({\rho}_2+\Lambda)-
(\rho-{\rho}_1-{\rho}_2)^2}\times\]
\[\left.\left.\ln\left[\frac {\sqrt{4({\rho}_1-\Lambda)({\rho}_2+\Lambda)-
(\rho-{\rho}_1-{\rho}_2)^2}-i(\rho-{\rho}_1-{\rho}_2)}
{2i\sqrt{-({\rho}_1-\Lambda)({\rho}_2+\Lambda)}}\right]\right\}\right\}- \]
\[\Theta[-\Im(\rho)]\left\{[
\ln(-{\rho}_1+\Lambda)-\ln(-{\rho}_1-\Lambda)]
[\ln(-{\rho}_2+\Lambda)-\ln(-{\rho}_2-\Lambda)]
\times\right.\]
\[\sqrt{4({\rho}_1-\Lambda)({\rho}_2-\Lambda)-
(\rho-{\rho}_1-{\rho}_2+2\Lambda)^2}\times\]
\[\ln\left[\frac {\sqrt{4({\rho}_1-\Lambda)({\rho}_2-\Lambda)-
(\rho-{\rho}_1-{\rho}_2+2\Lambda)^2}-i(\rho-{\rho}_1-{\rho}_2+2\Lambda)}
{2\sqrt{({\rho}_1-\Lambda)({\rho}_2-\Lambda)}}\right]+ \]
\[[\ln({\rho}_1+\Lambda)-\ln({\rho}_1-\Lambda)]
[\ln({\rho}_2+\Lambda)-\ln({\rho}_2-\Lambda)]\times\]
\[\sqrt{4({\rho}_1+\Lambda)({\rho}_2+\Lambda)-
(\rho-{\rho}_1-{\rho}_2-2\Lambda)^2}\times\]
\[\ln\left[\frac {\sqrt{4({\rho}_1+\Lambda)({\rho}_2+\Lambda)-
(\rho-{\rho}_1-{\rho}_2-2\Lambda)^2}-i(\rho-{\rho}_1-{\rho}_2-2\Lambda)}
{2\sqrt{({\rho}_1+\Lambda)({\rho}_2+\Lambda)}}\right]+ \]
\[[\ln({\rho}_1+\Lambda)-\ln({\rho}_1-\Lambda)]
[\ln(-{\rho}_2+\Lambda)-\ln(-{\rho}_2-\Lambda)]\times\]
\[\left\{\frac {i\pi} {2}\left[
\sqrt{4({\rho}_1-\Lambda)({\rho}_2+\Lambda)-
(\rho-{\rho}_1-{\rho}_2)^2}-i(\rho-{\rho}_1-{\rho}_2)\right]\right.+\]
\[\sqrt{4({\rho}_1-\Lambda)({\rho}_2+\Lambda)-
(\rho-{\rho}_1-{\rho}_2)^2}\times\]
\[\left.\ln\left[\frac {\sqrt{4({\rho}_1-\Lambda)({\rho}_2+\Lambda)-
(\rho-{\rho}_1-{\rho}_2)^2}-i(\rho-{\rho}_1-{\rho}_2)}
{2i\sqrt{-({\rho}_1-\Lambda)({\rho}_2+\Lambda)}}\right]\right\}+ \]
\[[\ln(-{\rho}_1+\Lambda)-\ln(-{\rho}_1-\Lambda)]
[\ln({\rho}_2+\Lambda)-\ln({\rho}_2-\Lambda)]\times\]
\[\left\{\frac {i\pi} {2}\left[
\sqrt{4({\rho}_1+\Lambda)({\rho}_2-\Lambda)-
(\rho-{\rho}_1-{\rho}_2)^2}-i(\rho-{\rho}_1-{\rho}_2)\right]\right.+\]
\[\sqrt{4({\rho}_1+\Lambda)({\rho}_2-\Lambda)-
(\rho-{\rho}_1-{\rho}_2)^2}\times\]
\[\left.\left.\ln\left[\frac {\sqrt{4({\rho}_1+\Lambda)({\rho}_2-\Lambda)-
(\rho-{\rho}_1-{\rho}_2)^2}-i(\rho-{\rho}_1-{\rho}_2)}
{2i\sqrt{-({\rho}_1+\Lambda)({\rho}_2-\Lambda)}}\right]\right\}\right\}-
\frac {i} {2}\times\]
\[\left\{[\ln(-{\rho}_1+\Lambda)-\ln(-{\rho}_1-\Lambda)]
[\ln(-{\rho}_2+\Lambda)-\ln(-{\rho}_2-\Lambda)]\right.\times\]
\[({\rho}_1-{\rho}_2)\left[\ln\left(i\sqrt{\frac {{\rho}_1+\Lambda}
{{\rho}_2+\Lambda}}\right)+
\ln\left(-i\sqrt{\frac {{\rho}_1-\Lambda}
{{\rho}_2-\Lambda}}\right)\right]+\]
\[[\ln({\rho}_1+\Lambda)-\ln({\rho}_1-\Lambda)]
[\ln({\rho}_2+\Lambda)-\ln({\rho}_2-\Lambda)]\times\]
\[({\rho}_1-{\rho}_2)\left[\ln\left(-i\sqrt{\frac {\Lambda-{\rho}_1}
{\Lambda-{\rho}_2}}\right)+
\ln\left(i\sqrt{\frac {\Lambda+{\rho}_1}
{\Lambda+{\rho}_2}}\right)\right]+\]
\[[\ln({\rho}_1+\Lambda)-\ln({\rho}_1-\Lambda)]
[\ln(-{\rho}_2+\Lambda)-\ln(-{\rho}_2-\Lambda)]\times\]
\[\left\{({\rho}_1-{\rho}_2)\left[\ln\left(\sqrt{\frac {\Lambda+{\rho}_1}
{\Lambda-{\rho}_2}}\right)+
\ln\left(\sqrt{\frac {\Lambda-{\rho}_1}
{\Lambda+{\rho}_2}}\right)\right]\right.+\]
\[\frac {({\rho}_1-{\rho}_2)} {2}\left[\ln(-{\rho}_1-{\rho}_2+\Lambda)-
\ln(-{\rho}_1-{\rho}_2-\Lambda)\right.-\]
\[\left.\ln({\rho}_1+{\rho}_2+\Lambda)+\ln({\rho}_1+{\rho}_2-\Lambda)\right]
+{\rho}_2\left[\ln(-{\rho}_1-{\rho}_2+\Lambda)\right.-\]
\[\left.\left.\ln(-{\rho}_1-{\rho}_2-\Lambda)\right]+{\rho}_1\left[
\ln({\rho}_1+{\rho}_2+\Lambda)-\ln({\rho}_1+{\rho}_2-\Lambda)\right]\right\}\]
\[[\ln(-{\rho}_1+\Lambda)-\ln(-{\rho}_1-\Lambda)]
[\ln({\rho}_2+\Lambda)-\ln({\rho}_2-\Lambda)]\times\]
\[\left\{({\rho}_1-{\rho}_2)\left[\ln\left(\sqrt{\frac {\Lambda-{\rho}_1}
{\Lambda+{\rho}_2}}\right)+
\ln\left(\sqrt{\frac {\Lambda+{\rho}_1}
{\Lambda-{\rho}_2}}\right)\right]\right.+\]
\[\frac {({\rho}_1-{\rho}_2)} {2}\left[\ln({\rho}_1+{\rho}_2+\Lambda)-
\ln({\rho}_1+{\rho}_2-\Lambda)\right.-\]
\[\left.\ln(-{\rho}_1-{\rho}_2+\Lambda)+\ln(-{\rho}_1-{\rho}_2-\Lambda)\right]
+{\rho}_1\left[\ln(-{\rho}_1-{\rho}_2+\Lambda)\right.-\]
\begin{equation}
\label{ep7.34}
\hspace{-6mm}
\left.\left.\left.\left.\ln(-{\rho}_1-{\rho}_2-\Lambda)\right]+{\rho}_2\left[
\ln({\rho}_1+{\rho}_2+\Lambda)-\ln({\rho}_1+{\rho}_2-\Lambda)\right]\right\}
\right\}\right\}\;d{\rho}_1\;d{\rho}_2
\end{equation}
Which defines an ultradistribution on the variables $\rho$ and $\Lambda$
for\\
$|\Im(\rho)|>\Im(\Lambda)>|\Im({\rho}_1)|+|\Im({\rho}_2)|$\\
Let {\textgoth{B}} be a vertical band contained in the complex
${\lambda}$-plane \textgoth{P}.
Integral (\ref{ep7.34}) is an analytic function of $\lambda$ defined in the
domain \textgoth{B}.
Moreover, it is bounded by a power of $|\rho\Lambda|$.
Then, according to the method of ref.\cite{tp7},
$H_{\lambda}(\rho,\Lambda)$ can be
analytically continued to other parts of \textgoth{P}.
Thus we define
\begin{equation}
\label{ep7.35}
H(\rho)=H^{(0)}(\rho,i0^+)
\end{equation}
\begin{equation}
\label{ep7.36}
H_{\lambda}(\rho,i0^+)=\sum\limits_{-m}^{\infty}
H^{(n)}(\rho,i0^+){\lambda}^n
\end{equation}
As in the other cases we define now
\begin{equation}
\label{ep7.37}
\{F\ast G\}(\rho)=H(\rho)
\end{equation}
as the convolution of two Lorentz invariant tempered ultradistributions.
The proof that $H(\rho)$ is a Tempered Ultradistribution is similar
to the one given in ref.\cite{tp3} for the one-dimensional case.
Starting with (\ref{ep7.34}) we can write:
\begin{equation}
\label{ep7.38}
H_{\lambda}(\rho,i0^+)=-\frac {1} {2\rho}
\iint\limits_{-\infty}^{\;\;\;\infty}f_{\lambda}({\rho}_1)g_{\lambda}({\rho}_2)
M(\rho.{\rho}_1,{\rho}_2)\;d{\rho}_1\;d{\rho}_2
\end{equation}
where $f_{\lambda}(\rho)$ and $g_{\lambda}(\rho)$ are defined by
Dirac's formula:
\begin{equation}
\label{ep7.39}
{\rho}^{\lambda}F_{\lambda}(\rho)=\frac {1} {2\pi i}
\int\limits_{-\infty}^{\infty}\frac {f_{\lambda}(t)} {t-\rho}\;dt\;\;;\;\;
{\rho}^{\lambda}G_{\lambda}(\rho)=\frac {1} {2\pi i}
\int\limits_{-\infty}^{\infty}\frac {g_{\lambda}(t)} {t-\rho}\;dt
\end{equation}
Let ${\hat{H}}_{\lambda}(x)$ be the Fourier antitransform of
$H_{\lambda}(\rho,i0^+)$. The according with (\ref{ep6.12}) to (\ref{ep6.17})
we can express $H^{(0)}(x)$ as a function of de Laurent developments
of ${\hat{f}}_{\lambda}(x)$ and  ${\hat{g}}_{\lambda}(x)$

\subsubsection*{Examples}

As an example of the use of (\ref{ep7.35}) we will evaluate the convolution
product of $\delta(\rho)$ with $\delta(\rho-{\mu}^2)$ with $\mu={\mu}_R+i{\mu}_I$
a complex number such that: ${\mu}_R^2>{\mu}_I^2$, ${\mu}_R{\mu}_I>0$.
Thus from (\ref{ep7.34}) we obtain:
\[H_0(\rho,\Lambda)=-i\pi[\ln(-{\mu}^2+\Lambda)-\ln(-{\mu}^2+\lambda)]
\left\{\frac {i(\rho-{\mu}^2)} {8{\pi}^2\rho}\left[
\ln\left(\frac {\rho-{\mu}^2} {\sqrt{\Lambda({\mu}^2+\Lambda)}}\right)+
\right.\right.\]
\[\left.\left.\ln\left(\frac {{\mu}^2-\rho} {\sqrt{-\Lambda({\mu}^2+\Lambda)}}\right)
\right]+\frac {{\mu}^2-\rho} {16\pi\rho}\right\}
-i\pi[\ln(-{\mu}^2+\Lambda)-\ln(-{\mu}^2+\lambda)]\;\;\times\]
\begin{equation}
\label{ep7.40}
\left\{\frac {-i{\mu}^2)} {8{\pi}^2\rho}\left[
\ln\left(\sqrt{\frac {\Lambda} {{\mu}^2+\Lambda}}\right)+
\ln\left(\sqrt{\frac {\Lambda} {\Lambda-{\mu}^2}}\right)
\right]-\frac {{\mu}^2} {16\pi\rho}\right\}
\end{equation}
Simplifying terms (\ref{ep7.47}) turns into:
\[H_0(\rho,\Lambda)=-i\pi[\ln(-{\mu}^2+\Lambda)-\ln(-{\mu}^2+\lambda)]
\left\{\frac {i(\rho-{\mu}^2)} {8{\pi}^2\rho}\left[
\ln\left(\rho-{\mu}^2\right)+
\right.\right.\]
\begin{equation}
\label{ep7.41}
\left.\left.\ln\left({\mu}^2-{\rho}\right)\right]+
\frac {i{\mu}^2} {8{\pi}^2\rho}\left[\ln({\mu}^2+\Lambda)+
\ln({\mu}^2-\Lambda)\right]\right\}
\end{equation}
Now, if
\[F_1(\mu,\Lambda)=\ln(-{\mu}^2+\Lambda)-\ln(-{\mu}^2-\Lambda)\]
then
\[F_1(\mu,i0^+)=2i\pi\;\;;\;\; {\mu}_R^2>{\mu}_I^2\;\;;\;\;{\mu}_R{\mu}_I>0\]
And, if
\[F_2(\mu,\Lambda)=\ln({\mu}^2+\Lambda)-\ln({\mu}^2-\Lambda)\]
then
\[F_2(\mu,i0^+)=0\;\;;\;\; {\mu}_R^2>{\mu}_I^2\;\;;\;\;{\mu}_R{\mu}_I>0\]
Using these results we obtain:
\begin{equation}
\label{ep7.42}
H(\rho)=\frac {i(\rho-{\mu}^2)} {4\rho}\left[
\ln(\rho-{\mu}^2)+\ln({\mu}^2-\rho)\right]+
\frac {i{\mu}^2} {2\rho}\ln({\mu}^2)
\end{equation}
As an example of the use of (\ref{ep6.17}) we will evaluate the convolution
product of two Dirac's delta: $\delta(\rho)\ast\delta(\rho)$. In this case we have:
\begin{equation}
\label{ep7.43}
F_{\lambda}(\rho)=-\frac {{\rho}^{\lambda-1}} {2\pi i}
\end{equation}
and as a consequence:
\begin{equation}
\label{ep7.44}
f_{\lambda}(\rho)=\frac {\sin(\pi\lambda)} {\pi} {\rho}_-^{\lambda-1}
\end{equation}
The Fourier antitransform of (\ref{ep7.44}) is:
\begin{equation}
\label{ep7.45}
{\hat{f}}_{\lambda}(x)
-\frac {2^{2\lambda}}{4{\pi}^3}\frac {\Gamma(1+\lambda)} {\Gamma(1-\lambda)}
\left[x_+^{-\lambda-1}-\cos(\pi\lambda)x_{-}^{-\lambda-1}\right]
\end{equation}
which can be written as:
\[{\hat{f}}_{\lambda}(x)
-\frac {2^{2\lambda}}{4{\pi}^3}\frac {\Gamma(1+\lambda)} {\Gamma(1-\lambda)}
\left[\frac {\cos(\pi\lambda)-1} {\lambda}\delta(x)+
x_+^{-1}-\cos(\pi\lambda)x_{-}^{-1}\right.\;\;+\]
\begin{equation}
\label{ep7.46}
\left.S_+^{-\lambda-1}-\cos(\pi\lambda)S_{-}^{-\lambda-1}\right]
\end{equation}
Thus we have:
\[{\hat{f}}_{\lambda}^2(x)
-\frac {2^{4\lambda}}{16{\pi}^6}\frac {{\Gamma}^2(1+\lambda)} {{\Gamma}^2(1-\lambda)}
\left\{\frac {(\cos(\pi\lambda)-1)^2} {{\lambda}^2}{\delta}^2(x)+
x_+^{-2}+{\cos}^2(\pi\lambda)x_{-}^{-2}\right.\;\;+\]
\[\left[S_+^{-\lambda-1}-\cos(\pi\lambda)S_{-}^{-\lambda-1}\right]^2+
2[x_+^{-1}-\cos(\pi\lambda)x_{-}^{-1}]
[S_+^{-\lambda-1}-\cos(\pi\lambda)S_{-}^{-\lambda-1}]+\]
\begin{equation}
\label{ep7.47}
\left.2\left[\frac {\cos(\pi\lambda)-1} {\lambda}\delta(x)\right]
\left[x_+^{-1}-\cos(\pi\lambda)x_{-}^{-1}+
S_+^{-\lambda-1}-\cos(\pi\lambda)S_{-}^{-\lambda-1}\right]\right\}
\end{equation}
From (\ref{ep7.47}) we obtain:
\begin{equation}
\label{ep7.48}
\lim_{\lambda\rightarrow 0}{\hat{f}}_{\lambda}^2(x)=
\frac {4} {(2\pi)^6}x^{-2}
\end{equation}
and taking into account that:
\begin{equation}
\label{ep7.49}
{\cal F}\{x^{-2}\}=\frac {{\pi}^3} {2} Sgn(\rho)
\end{equation}
we obtain
\begin{equation}
\label{ep7.50}
\delta(\rho)\ast\delta(\rho)=\frac {\pi} {2} Sgn(\rho)
\end{equation}

\newpage

\section{Discussion}

In a earlier paper \cite{tp3} we have shown the existence of the convolution
of two one-dimensional tempered ultradistributions. In other paper
ref.\cite{tq1} we have
extended these procedure to n-dimensional space. In four-dimensional
space we have given an expression for the convolution of two tempered
ultradistributions even in the variables $k^0$ and $\rho$.
In this paper we obtain a expression for the convolution of two Lorentz
invariant tempered ultradistributions in both, Euclidean and Minkowskian space.
In an intermediate step of deduction we obtain the generalization
to the Minkowskian space of the dimensional regularization in
configuration space (ref.\cite{tp12})

When we use the perturbative development in Quantum Field Theory, we
have to deal with products of distributions in configuration space,
or else, with convolutions in the Fourier transformed p-space.
Unfortunately, products or convolutions ( of distributions ) are
in general ill-defined quantities. However, in physical applications
one introduces some ``regularization'' scheme, which allows us to
give sense to divergent integrals. Among these procedures we would
like to mention the dimensional regularization method ( ref.
\cite{tp14,tp15} ). Essentially, the method consists in the
separation of the volume element ( $d^{\nu}p$ ) into an angular
factor ( $d\Omega$ ) and a radial factor ( $p^{\nu-1}dp$ ).
First the angular integration is carried out and then the number
of dimensions $\nu$ is taken as a free parameter. It can be adjusted
to give a convergent integral, which is an analytic function of
$\nu$.

Our formula (7.34) is similar to the expression one obtains with
dimensional regularization. However, the parameter $\lambda$ is
completely independents of any dimensional interpretation.

All ultradistributions provide integrands ( in (7.34) ) that are
analytic functions along the integration path. The parameter
$\lambda$ permits us to control the possible tempered asymptotic
behavior ( cf. eq. (3.9) ). The existence of a region of
analyticity in $\lambda$, and a subsequent continuation to
the point of interest ( ref. \cite{tp3} ), defines the convolution
product.

The properties described below
show that tempered ultradistributions provide an
appropriate framework for applications to physics. Furthermore,
they can ``absorb'' arbitrary pseudo-polynomials, thanks to eq. (3.10).
A property that is interesting for renormalization theory.
 For this reason and also for the benefit
 of the reader
we began this paper with a summary of the main characteristics
of n-dimensional tempered ultradistributions and their Fourier
transformed
distributions of the exponential type.

As a final remark we would like to point out that our formula for convolutions
is a definition and not a regularization method.


\newpage

\begin{thebibliography}{99}

\bibitem{tp1} J. Sebastiao e Silva : Math. Ann. {\bf 136},
38 (1958).
\bibitem{tp2} M. Hasumi: T$\rm{\hat{o}}$hoku Math. J. {\bf 13},
94 (1961).
\bibitem{tp3} C. G. Bollini, T. Escobar and M. C. Rocca :
Int. J. of Theor. Phys. {\bf 38}, 2315 (1999).
\bibitem{tq1} C. G Bollini and M.C. Rocca : ``Convolution of n-dimensional
tempered ultradistributions and Field Theory'': hep-th /0309271. To be
published in Int. J. of Theor. Phys.
\bibitem{tp4} I. M. Gel'fand and N. Ya. Vilenkin : ``Generalized
Functions'' {\bf Vol. 4}. Academic Press (1964).
\bibitem{tp5} I. M. Gel'fand and G. E. Shilov : ``Generalized
Functions'' {\bf Vol. 2}. Academic Press (1968).
\bibitem{tp6} L. Schwartz : ``Th\'eorie des distributions''.
Hermann, Paris (1966).
\bibitem{tp7} I. M. Gel'fand and G. E. Shilov : ``Generalized
Functions'' {\bf Vol. 1} Ch.1, section 3. Academic Press (1964).
\bibitem{tp8} L. S. Gradshtein  and I. M. Ryzhik : ``Table of Integrals, Series, and
Products''. Sixth edition.  Academic Press (2000).
\bibitem{tp9} G. N. Watson : ``A Treatise on the Theory of Bessel Functions''
Second edition. Cambridge University Press (1995)
\bibitem{tp10} I. M. Gel'fand and G. E. Shilov : ``Generalized
Functions'' {\bf Vol. 1}. Academic Press (1964).
\bibitem{tp11} A. Erdelyi : ``Bateman Manuscript Project'' Vol 1 :
Higher Transcendental Functions. McGraw-Hill Book Company,
Inc. (1953).
\bibitem{tp12} C. G. Bollini and J.J Giambiagi :
Phys. Rev. {\bf D 53}, 5761 (1996).
\bibitem{tp13} Y. A. Brychkov and A. P. Prudnikov : ``Integral Transforms
of Generalized Functions''. Gordon and Breach Science Publishers. (1989).
\bibitem{tp14} C. G. Bollini and J. J. Giambiagi: Phys. Lett.
{\bf B 40}, 566 (1972). Il Nuovo Cim. {\bf B 12}, 20 (1972).
\bibitem{tp15} G.'t Hooft and M. Veltman: Nucl. Phys. {\bf B 44},
189 (1972).

\end{thebibliography}
\end{document}